\documentclass[12pt]{iopart}

\usepackage[a4paper]{geometry}
\usepackage{color}
\usepackage{graphicx,amssymb,amstext}
\usepackage{framed}
\usepackage{epigraph}
\usepackage{verbatim}
\usepackage{float}
\usepackage{calligra}
\usepackage{subcaption}

\usepackage{hyperref}

\newcommand{\refeq}[1]{Eq.(\ref{#1})}
\newcommand{\reffig}[1]{Fig.\;\ref{#1}}
\newcommand{\refsec}[1]{\S\ref{#1}}

\newcommand{\infinity}{\infty}
\newcommand{\abs}[1]{\left|#1\right|}
\newcommand{\integers}{\mathbb{Z}}
\newcommand{\complex}{\mathbb{C}}
\newcommand{\real}{\mathbb{R}}

\newcommand{\fract}[2]{{\textstyle\frac{#1}{#2}}}

\newcommand{\half}{\frac{1}{2}}
\newcommand{\halft}{\fract{1}{2}}

\begin{document}

\textheight=22cm

\title[Breaking integrability at the boundary: the sine-Gordon model]%
{Breaking integrability at the boundary: the sine-Gordon model
with Robin boundary conditions}
\author{Robert Arthur$^1$, Patrick Dorey$^1$, Robert Parini$^2$}

\address{$^1$ Department of Mathematical Sciences, Durham University, UK}

\address{$^2$ Department of Mathematics, University of York, UK}

\begin{abstract}
	We explore boundary scattering in the
sine-Gordon model with a non-integrable family of Robin boundary
conditions.
The soliton content of the field after collision is analysed using a
numerical implementation of the direct scattering problem associated
with the inverse scattering method.  We find that
an antikink may be reflected into various combinations of
an antikink, a kink, and one or more breathers,
depending on the
values of the initial antikink velocity and a parameter associated
with the boundary condition.  In addition
we observe regions with an intricate resonance structure arising from the
creation of an intermediate breather whose recollision with the boundary
is highly dependent on the breather phase.  
\end{abstract}


\section{Introduction}

The sine-Gordon equation has attracted considerable attention
over the years, partly for its application in physical contexts, 
but also as a model of an integrable equation admitting 
topologically non-trivial soliton solutions.
The (rescaled) equation on the full line
describes a single scalar field $u(x,t)$ satisfying
\begin{equation}
	u_{tt} - u_{xx} + \sin(u) = 0 \label{SG}
\end{equation}
and has vacua with the field taking the values ${u(x,t)=2\pi n}$, 
${n\in\integers}$. The particle-like excitations of the theory
are solitons (kinks or antikinks) of mass $8$ which interpolate
between neighbouring vacua, and
breathers with angular frequency $\omega$, $0<\omega<1$,
and mass $16\sqrt{1-\omega^2}$,
which are bound states of
kinks and antikinks.  Integrability constrains these excitations to
scatter
in a remarkably simple way, preserving their velocities and shapes
while undergoing a phase shift.

This simplicity extends to 
the half-line theory ($x<0$) provided a suitable boundary condition is
imposed. 
Following earlier work covering various special cases
(see for example \cite{Bikbaev1992a}), 
a consideration of the low-lying conserved charges led Ghoshal
and Zamolodchikov 
\cite{Ghoshal1993}  
to propose that
the most general boundary condition consistent with
integrability and arising from a boundary action without a kinetic
term or any additional degrees of freedom should be
\begin{equation}
	\left. \left[
	u_x + 4K \sin\left(\frac{u-\widehat u}{2}\right)
	\right] \right|_{x=0}
	= 0,
	\label{integrableBC}
\end{equation}
with $K$, $\widehat u$ $\in\real$. 
The existence of an infinite set of conserved quantities for these
boundary conditions was established in \cite{MacIntyre1994}, and, as
will be discussed further in \refsec{conclusions}, their special role
within the Fokas (or unified) method was elucidated in
\cite{Fokas2003}. 
If an antikink is sent towards such a boundary then depending upon its
initial velocity $v_0$ and the boundary parameters $\widehat u$ and $K$ it
will return as either an antikink or a kink but without any loss of
energy, having only experienced a phase shift \cite{Saleur1994}.

A much wider variety of final states is possible if one instead
considers non-integrable boundaries.  A kink (sometimes referred to as
a fluxon) colliding with the `magnetic field' boundary condition
$u_x(0,t)=\beta$ was found in \cite{Olsen1981} either to collapse into
radiation, to produce one or more kinks or antikinks, or to produce a
breather depending on the initial kink velocity and $\beta$.  This
boundary condition arises as a model of an external magnetic field of
magnitude $\beta$ applied to a long, narrow Josephson junction. 

In this paper we will consider instead a field $u(x,t)$ satisfying the 
sine-Gordon equation (\ref{SG})
for $x<0$ with the homogeneous Robin boundary condition
\begin{equation}
	u_x + 2ku = 0 
\label{robin}
\end{equation}
imposed at $x=0$.
This boundary may be derived as the linearisation of
\refeq{integrableBC} with $\widehat u=0$ and even though it is non-integrable
for general $k$, it possesses integrable limits.  For
$k\rightarrow\infinity$ the boundary becomes Dirichlet, $u(0,t)=0$, and for
$k=0$ it becomes Neumann, $u_x(0,t)=0$.  This allows us to consider `close'
to integrable situations when $k$ is very small or large.  To see how
the boundary interpolates between the two integrable limits we will
send an antikink with initial velocity $v_0$ towards the
boundary, and analyse the behaviour of the field at large times after the
collision.

Notice that even away from the integrable limits, any reflected
excitations will ultimately be far from the boundary at which
integrability is broken, and thus should be describable in terms of the 
well-understood set of excitations of the integrable full-line theory. 
This makes the problem somewhat cleaner than boundary
scattering in models such as the $\phi^4$ theory, recently studied in
\cite{Dorey2015}, where integrability is absent in the
bulk, and it also means that we will be able to
unravel the soliton and breather content of the field
after the collision 
by numerically determining the scattering data associated
with the corresponding full-line inverse scattering transform.

\section{Numerical method}

On the full line, the sine-Gordon equation is integrable and
the initial value problem for
asymptotically decaying initial conditions can be solved by the
inverse scattering method
\cite{Takhtadzhyan1974,Ablowitz1981,Forest82}.  
This involves considering the pair of linear eigenvalue problems
\begin{eqnarray}
	\psi_x & = V(u,u_x,u_t;\lambda) \psi \label{AKNS}\\
	\psi_t & = U(u,u_x,u_t;\lambda) \psi \label{AKNSt}
	\qquad
	\lambda \in \complex,
\end{eqnarray}
where the eigenfunction $\psi$ is a $2\times1$ column vector and $V$
and $U$ are $2\times2$ matrix-valued functions such that the
compatibility condition of Eqs.\,(\ref{AKNS}) and (\ref{AKNSt}),
$\psi_{xt}=\psi_{tx}$, implies the sine-Gordon equation.  To solve the 
equation given some initial functions $u$ and $u_t$ one would first use
\refeq{AKNS} to obtain the scattering data, a part of which is the set
of bound state eigenvalues $\{\lambda_n\}$ in the upper half plane
$\text{Im}[\lambda]>0$ at which \refeq{AKNS} has a
solution decaying at both plus and minus infinity;
then perform the time
evolution of the scattering data using \refeq{AKNSt}; and finally
reconstruct the field at the later time from the time-evolved
scattering data.  However, for our purposes we only need that
$\{\lambda_n\}$ encodes the velocities and frequencies of the soliton 
and breather
content of the field in a very simple way
\cite{Takhtadzhyan1974,Forest82}. Eigenvalues occur either on the
positive imaginary axis corresponding to kinks or antikinks, 
or in symmetrically-placed pairs $(\lambda_n,-\lambda_n^*)$ 
corresponding to breathers. Their velocities  and (in the case of
breathers) frequencies
are 
\begin{equation}
 v = \frac{1-16\abs{\lambda_n}^2}{1+16\abs{\lambda_n}^2}\,,
 \qquad 
 \omega = \frac{\text{Re[$\lambda_n$]}}{\abs{\lambda_n}}\,,
\label{kinematics}
\end{equation}
while their energies are
\begin{equation}
 E_{\rm soliton}=
\frac{1}{\abs{\lambda_n}}+16\abs{\lambda_n}\,,
 \quad 
 E_{\rm breather}=
2\,\text{Im[$\lambda_n$]}\left(\frac{1}{\abs{\lambda_n}^2}+16\right).
\label{energies}
\end{equation}
On the half line with a generic Robin boundary condition 
at $x=0$, integrability is lost and the inverse scattering
method cannot be used in any straightforward way. However if the
boundary partial differential equation is used
numerically, to evolve
an initial right-moving antikink far 
enough past the time of
its collision with the boundary, we would expect to reach a
stage when all excitations produced in the collision 
have departed from the boundary region and are moving back to the left,
away from the boundary. The subsequent evolution of these excitations
will then be the same, to a good approximation, as that of the
corresponding excitations on a full line.
The reflected field on the portion of the half-line containing these
excitations can
then be smoothly extended
to a solution on the full line, and the soliton
and breather content 
can be extracted by solving the full-line direct scattering
problem as just described.
This approach would fail, or at least miss some of the story, if
an infinitely long-lived boundary excitation -- a form of `boundary breather'
-- were to form during the collision. 
This possibility seems unlikely
given the loss of integrability at the boundary and we saw no sign of
it in our results, so we will disregard it in the following.
Nevertheless, a more detailed analytical and numerical
investigation of the timescales over which energy leaks away from the
boundary after a collision would be an interesting avenue for further
work.

\subsection{Time evolution}
\label{TEvolve}

We evolved the initial antikink
forward in time using a simple Euler finite difference scheme.
The initial profile was 
\begin{equation}
u(x,0) = 4\arctan\left(e^{-\gamma(v_0)(x-x_0)}\right),\quad
\gamma(v_0)=(1-v_0^2)^{-1/2}\,,
	\label{antikink}
\end{equation}
where  ${v_0 > 0}$ is the initial velocity and $x_0<0$
the initial position. We set $|x_0|=30$, to ensure that
the initial configuration satisfied
the Robin boundary condition at $x=0$ 
to a good approximation, with $v_0$ effectively
the initial velocity of an antikink arriving from minus infinity.
(Since the discrete $u\rightarrow-u$ symmetry of the bulk
equation is respected by the Robin boundary, nothing new would be
gained by instead considering kink collisions.)
For most of the numerical work the space and time
steps were $dx=0.025$ and $dt=0.02$, but in situations with higher
sensitivity to errors a finer grid was used: $dx=0.0025$ and
$dt=0.002$ for figures
\ref{antikinkTrajectory}, 
\ref{breatherTrajectory},
\ref{2piantikinkTrajectory},
\ref{spaceTime},
\ref{kinematicsHigh},
\ref{kinematicsMess},
\ref{spaceTimeMess}
and \ref{spaceTimePhase}.
During time
evolution the left hand boundary $x=x_L$, at which $u=2\pi$, was dynamically 
extended so that
anything produced from the collision with the Robin boundary never reached it.
Effectively, this implemented the
boundary condition $u \rightarrow 2\pi$ as $x \rightarrow
-\infinity$.

\begin{figure}
\centering
\begin{subfigure}{0.5\textwidth}
  \centering
  \hspace* {-20pt}
  \includegraphics[width=0.98\linewidth]{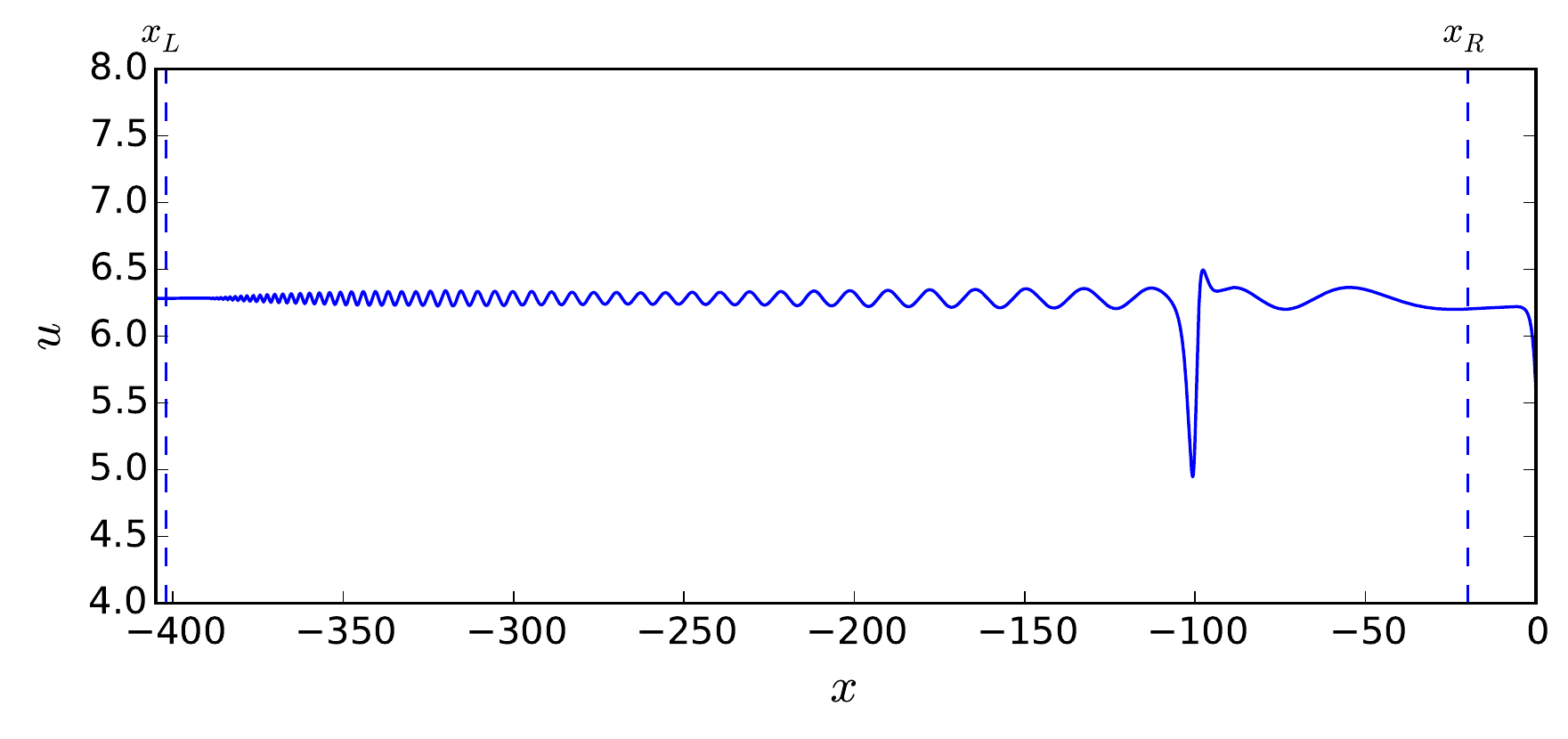}
  \caption{}
  \label{scatteredField}
\end{subfigure}%
\begin{subfigure}{0.5\textwidth}
  \centering
  \includegraphics[width=0.96\linewidth]{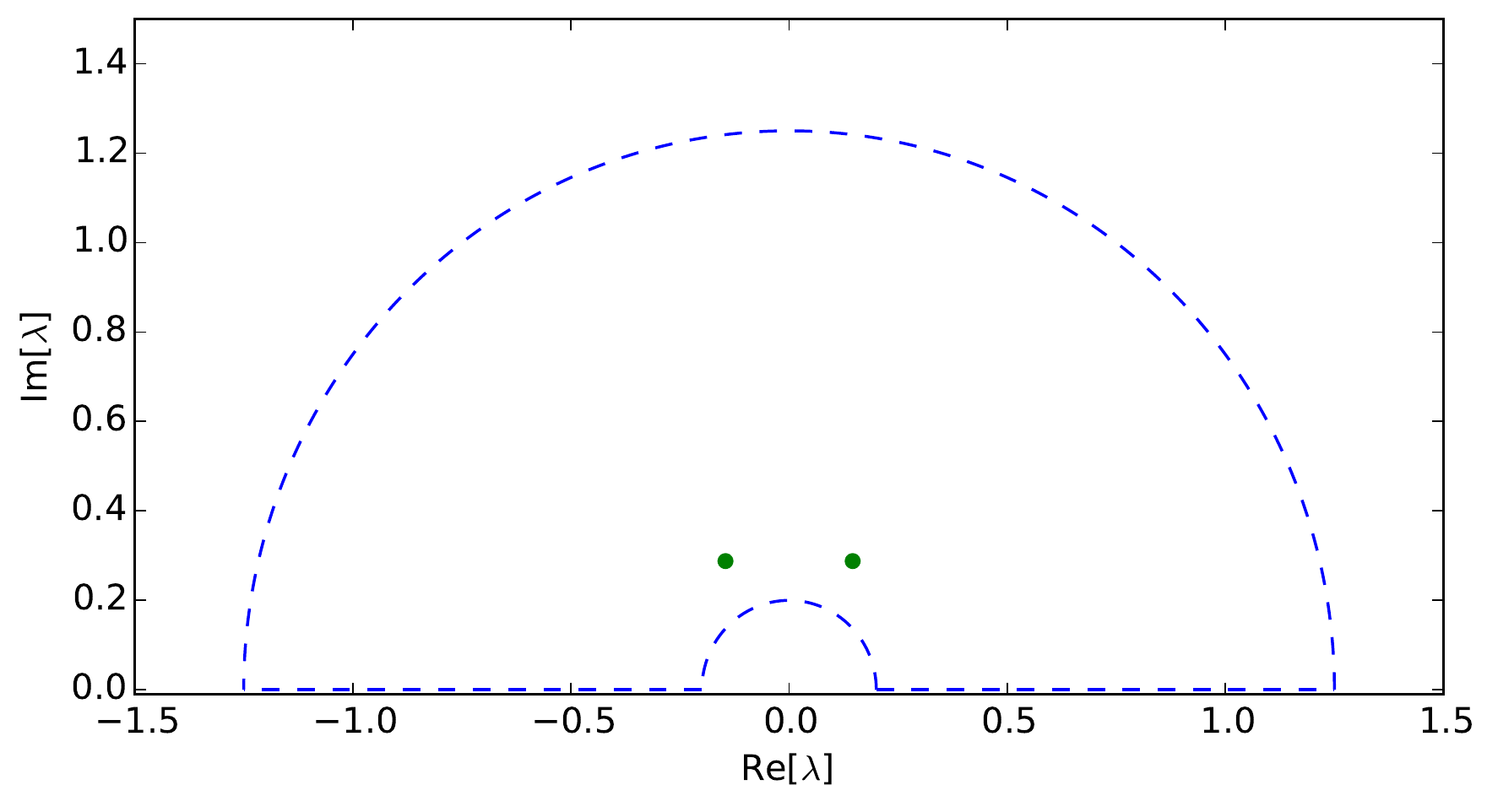}
  \caption{}
  \label{eigenvalues}
\end{subfigure}
\caption{\small
(a) The scattered field after time evolution with initial soliton
velocity $v_0=0.895$ and boundary parameter $k=0.055$.  (b) The
bound state eigenvalues, $\pm 0.146 + 0.288 i$, derived
from the portion of the field between $x_L$ and $x_R$ in (a).  They
correspond to a breather with frequency $0.45$ and velocity $-0.25$.
The dashed curve shows the initial integration contour.}

\label{example}
\end{figure}

In order to be able to extract the scattering data the
reflected field must have reached a vacuum value, $u=2\pi n$,
$n\in\integers$ at the right-hand boundary as well.
However, an important feature of our Robin boundary 
condition is that $u=0$ is the only zero energy state
for the field, and in cases where the
topological charge of the field \footnote{We will loosely define the
topological charge as the number of kinks minus the number of
antikinks because the value of the field at $x=0$ will not in general
be an integer multiple of $2\pi$.} is changed during the
scattering there will be a slight deformation of the field close
to the boundary (as in \reffig{scatteredField}) to satisfy the Robin
boundary condition.  Because of this we will have to take the right
hand boundary for the purposes of the direct scattering problem to be
some $x_R < 0$ (typically $x_R=-20$), and wait until
all excitations of interest are in the region $x<x_R$.

To achieve this we exit the time evolution if the field and its
derivatives at $x=x_R$ are sufficiently close to a vacuum and the
total available energy\,\footnote{The `available energy' is the energy
in the field $x_R<x<0$ in addition to the energy due to the boundary,
$k u(x=0)^2$, subtracting the energy of the minimum-energy
configuration satisfying the boundary condition.  This
will be zero if after the collision $u(x=0)=0$;
otherwise it can be extracted from a hypothetical static
antikink placed near the boundary such that the boundary
condition is satisfied.} in the region $x_R<x<0$ is less than $1$,
well below the mass of a single kink or antikink. Or failing that if a time
of $1000 + \abs{x_0}/v_0$ has elapsed.
This ensures that any excitations with significant energy have been emitted
from the boundary and allows us to embed the segment $x_L\leq x\leq
x_R$ into 
the full line and analyse its field content via the direct scattering
problem.

\subsection{Direct scattering problem}

After time evolution we consider the linear eigenvalue problem,
\refeq{AKNS}, with \cite{Takhtadzhyan1974}
\begin{equation}
	V(u,u_x,u_t;\lambda) = 
	\left(
	\begin{array}{cc}
		-\frac{i(u_x+u_t)}{4} & \lambda - \frac{e^{-iu}}{16\lambda} \\
		\frac{e^{iu}}{16\lambda} - \lambda & \frac{i(u_x+u_t)}{4}
	\end{array}
	\right).
	\label{potential}
\end{equation}
If the field tends to the vacuum, $u \rightarrow 2n\pi, n \in \mathbb{Z}$ and $u_{t}$, $u_{x} \rightarrow 0$, 
as $|x|\to\infty$ then for $\text{Im}[\lambda]>0$
two solutions $\psi_+$ and $\psi_-$ to
\refeq{AKNS} can be defined at any fixed time $t$ via the asymptotics
\begin{eqnarray}
\psi_-(x)
&\sim&
	\left(\begin{array}{c} 1\\ \!-i\! \end{array}\right) 
e^{-i(\lambda - 1/16\lambda)x} 
\qquad \text{as}~ x\to -\infty\,,
\label{psiminus}
\\[3pt]
\psi_+(x) &\sim& 
	\,\left(\begin{array}{c} 1\\i \end{array}\right)
	e^{\,i(\lambda - 1/16\lambda)x}
\qquad~\, \text{as}~ x\to +\infty\,.
\label{psiplus}
\end{eqnarray}
Note that $\psi_-(x)$ decays as $x\to-\infty$, and
$\psi_+(x)$ decays as $x\to+\infty$.

Since $x_L$ is defined as a point to the left of anything generated by the collision, $u(x_L)=2 \pi$ and $u_t(x_L)=u_x(x_L)=0$.  
At $x_R$ we assume, based on the conditions for ending the time evolution discussed in \refsec{TEvolve}, that the field and its derivatives are sufficiently close to the vacuum, as illustrated in \reffig{scatteredField}.
For the purposes of our computation we therefore effectively identify the points $-\infty$ and $+\infty$ as they relate to the direct scattering problem on the full line with $x_L$ and $x_R$ respectively.

We can then solve \refeq{AKNS} for any given $\lambda$ as an initial value problem for $\psi_-(x)$ from $x=x_L$ to $x=x_R$ with the initial condition $\psi_-(x_L)$ defined by the asymptotic form \refeq{psiminus}.
If $\lambda$ is one of the bound state eigenvalues $\lambda_n$, then $\psi_-(x)
\propto \psi_+(x)$, and so determining the bound state eigenvalues for a 
given reflected field amounts to finding the zeros of the Wronskian
\begin{equation}
	W(\lambda) = \text{Det}\left[ \psi_-(x=x_R), \psi_+(x=x_R) \right],
\end{equation}
which is a complex analytic function in the Im$[\lambda]>0$ region
\cite{Takhtadzhyan1974}. 
The value of $\psi_-(x_R)$ is the result of solving \refeq{AKNS} over the interval $x_L<x<x_R$ while $\psi_+(x_R)$ is given by the asymptotic form \refeq{psiplus}.

To find the zeros of $W$, we used the QZ-40
algorithm proposed in \cite{Dellnitz2002} with the numerical
integration performed using the Romberg algorithm as implemented in SciPy
\cite{scipy}.  QZ-40 employs the argument principle to find all the
simple roots of a complex analytic function within a given initial
contour.
This initial contour was chosen by first noting that any excitation in
the final state
must have $v < 0$ so by \refeq{kinematics}
$\abs{\lambda_n}>0.25$.  For kinks and antikinks, conservation of
energy implies that the reflected soliton speed $\abs{v} \leq v_0$
which gives $\abs{\lambda_n} < (1+v_0)/[4(1-v_0)]$.  However,
breathers can have a higher frequency (lower mass) so we cannot give a
bound on their speed.  To mitigate this we always took the outer
radius at least equal to $1.25$, meaning that we always detected 
breathers with speeds below $0.923$.
Some high frequency breathers may thereby go undetected, but
these will be of very low energies and therefore
largely insignificant to the overall reflection process.

Once a root $\lambda_n$ of the Wronskian has been found, the speed and 
frequency of the corresponding excitation can be read 
from \refeq{kinematics}, as illustrated in \reffig{eigenvalues}.  

\subsection{A test case}

\begin{figure}
	\centering
	\includegraphics[width = 0.75\linewidth]{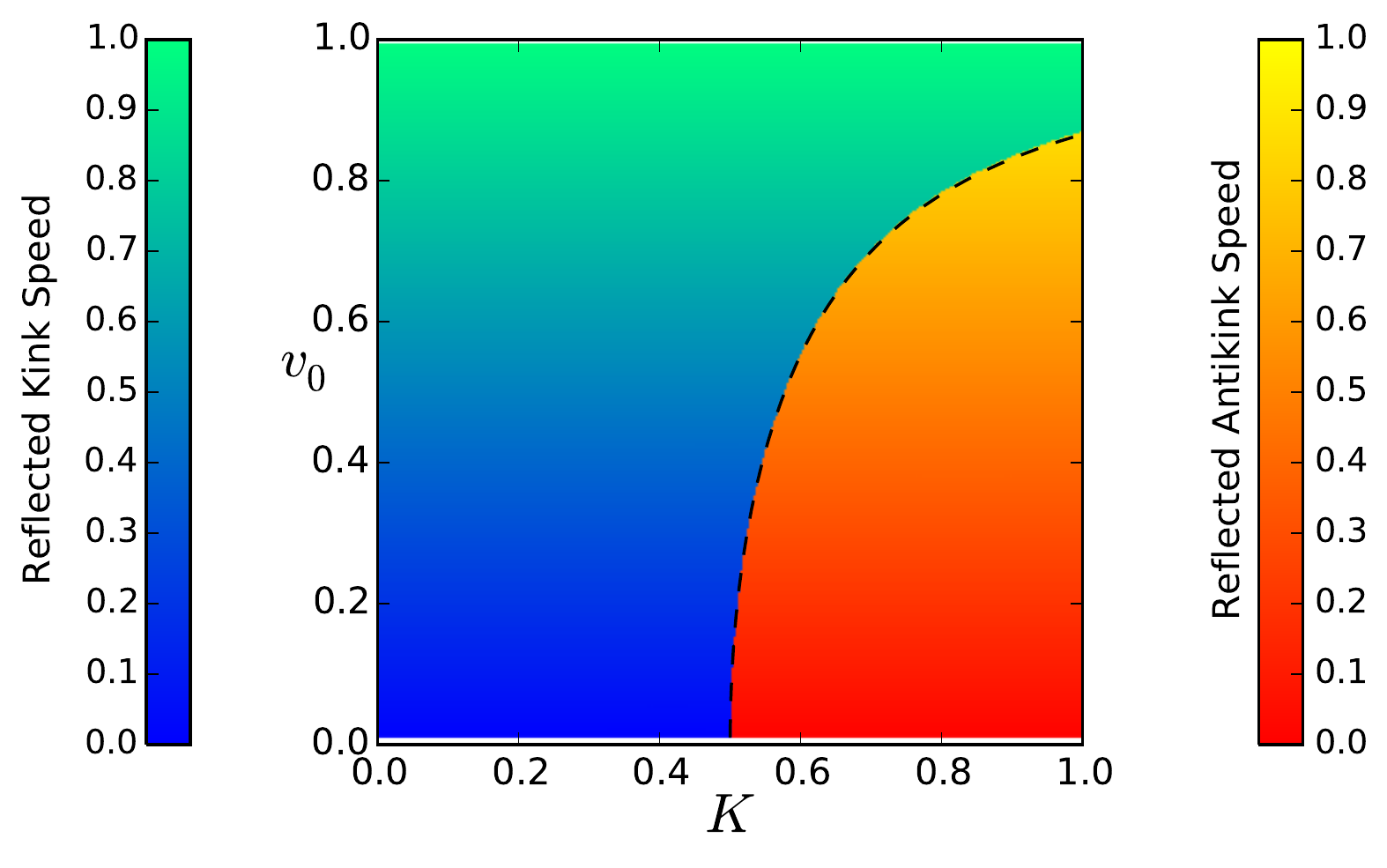}
\vskip -8pt
	\caption{\small
Numerical results for the
final kink/antikink speed for an initial antikink with speed $v_0$
hitting the integrable boundary (\ref{integrableBC}) with
boundary parameter $K$ and $\widehat u = 0$.
		The dashed line, $2K \sqrt{1-v_0^2} = 1$, is the
analytically-determined boundary between where a kink and antikink is
returned from the boundary collision \cite{Saleur1994}.  Precisely on
this line the incoming antikink should, theoretically, be infinitely
phase shifted.  }
	\label{integrableEigenvalue}
\end{figure}

As a simple test of our method we first collided an antikink
with the integrable boundary condition, \refeq{integrableBC}, with
$\widehat u = 0$.  The measured velocity of the reflected kink or antikink
after the collision is shown in \reffig{integrableEigenvalue}.  Over
the range of $v_0$ and $K$ shown in the figure the
maximum difference between the theoretical and measured final speed
was $0.0014$.  \reffig{integrableEigenvalue} also shows a
very good match between the observed and theoretical boundary between
the regions where the antikink is reflected into a kink or an antikink.

\section{The Robin boundary: analytic properties}
Before discussing our numerical results 
we derive some general properties of the theory with Robin boundary
conditions that can be obtained analytically. Throughout this
section and the next we assume that $k\ge 0$.

\subsection{Vacua and vacuum energies}
\label{Vac}
The Neumann boundary admits infinitely-many degenerate vacua, matching
the bulk: $u(x)=2\pi n$, $n\in\integers$. By contrast the
theory with the homogeneous Dirichlet condition
$u|_{x=0}=0$ has only one vacuum, $u(x)=0$. The Robin boundary
\refeq{robin}
with $k>0$ interpolates between these two situations as follows.

\begin{figure}
\centering
\hspace{0.3cm}
\includegraphics[width = 0.9\linewidth]{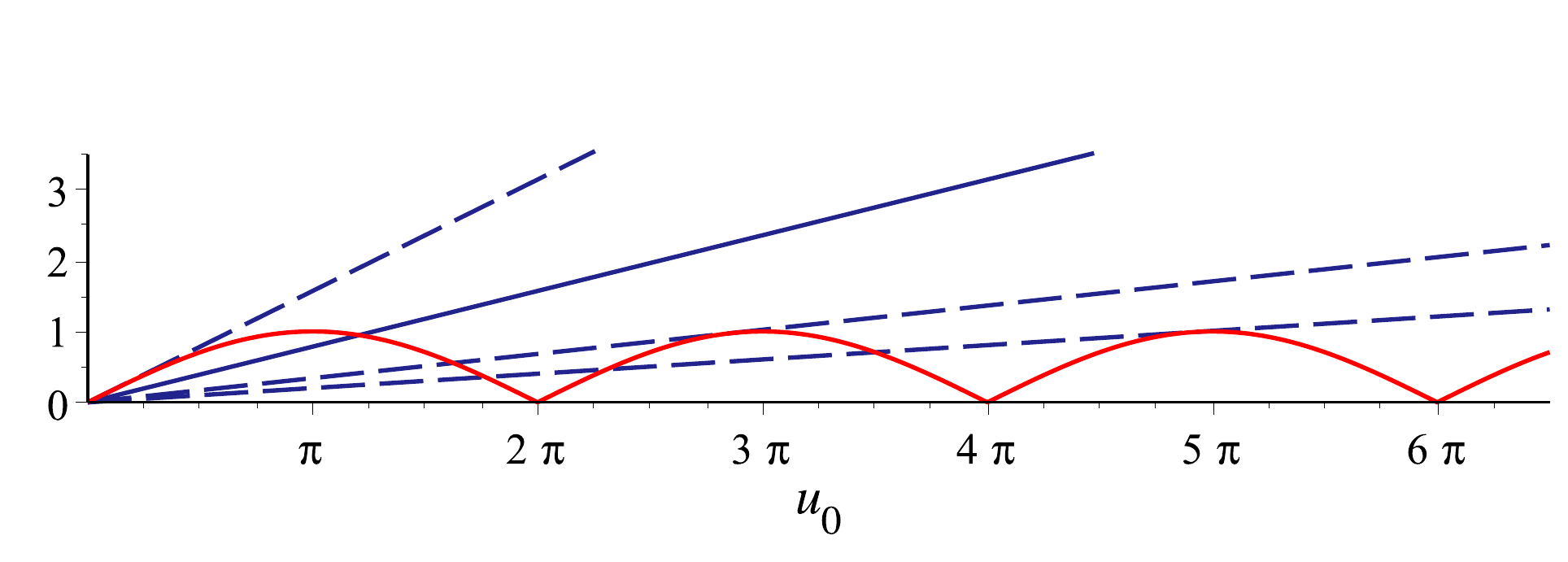}
\caption{\small
The graphical solution of \refeq{vaceq} for $k=0.25$ (solid line), 
and for the first three critical values of $k$ (dashed lines).}
	\label{vacua}
\end{figure}

As $x\to-\infty$, $u(x)$ must tend to one of the bulk vacua:
$u(x)\to 2\pi n$. For $-\infty<x<0$ it
must if static satisfy the relevant Bogomolnyi equation, which is
\begin{equation}
u_x=2\varepsilon\sin(u/2) 
\label{bog}
\end{equation}
where $\varepsilon=(-1)^n$.
Taking the limit of \refeq{bog} as $x\to 0$, the Robin
boundary condition $u_x|_{x=0}+2ku|_{x=0}=0$ can be rewritten as
$ku_0= -\varepsilon\sin(u_0/2)$,
where $u_0=u(0)$. Supposing 
$n$ to be positive for now, we have $2\pi (n{-}1)<u_0<2\pi n$,
so
$\text{sign}(\sin(u_0/2))=-\varepsilon$ and the boundary condition
to be satisfied is
\begin{equation}
ku_0= |\sin(u_0/2)|\,.
\label{vaceq}
\end{equation}
The graphical solution of this equation is illustrated in
\reffig{vacua}. As $k$ decreases from $+\infty$ (Dirichlet) towards $0$
(Neumann), the number of nonnegative
static solutions to the boundary problem
jumps from $1$ to $2$ at $k=0.5$, then to $4$ and so
on.  Transitions occur at the critical values $k=k_j$
where $k_j=|\frac{1}{2}\cos(u_0^{(j)}/2)|$ and the numbers
$u_0^{(j)}\ge 0$ solve 
$u_0^{(j)}=2\tan(u_0^{(j)}/2)$. For each nontrivial positive solution
there is a corresponding negative solution, so the \textit{total} number of
static solutions jumps from $1$ to $3$ to $7$ and so on as $k$
decreases.

The total energy 
$E=\int_{-\infty}^0\half(u_x)^2+(1{-}\cos u)\,dx + ku_0^2$
of any of these static solutions can be computed by recasting $E$ in
Bogomolnyi form as
\begin{eqnarray}
E&=&\halft\int_{-\infty}^0 \left(u_x-2\varepsilon\sin(u/2)\right)^2dx
-\varepsilon\left[4\cos(u/2)\right]_{-\infty}^0+ku_0^2 \nonumber\\
&=&
4-4\varepsilon\cos(u_0/2)+ku_0^2\,.
\label{staticEnergy}
\end{eqnarray}
where ${\varepsilon=-\text{sign}(\sin(u_0/2))}$. This function is
illustrated in \reffig{vacuaEnergy} below; as further
explained in the caption,
it has discontinuities whenever $u_0$ is an integer multiple of $2\pi$
and $\varepsilon$ changes sign.
Note that $\frac{dE}{du_0}=2\varepsilon\sin(u_0/2)+2ku_0= 
-2|\sin(u_0/2)|+2ku_0$, so $E$ is stationary as a function of $u_0$
exactly when the boundary condition \refeq{vaceq} holds, as has to be
the case. Furthermore, as is clear from \reffig{vacua}, for $n\ge 2$
$\frac{dE}{du_0}$
is negative in the interval between any two of its zeros which both lie in
an interval $2\pi (n{-}1)<u_0<2\pi n$ and positive outside it, 
so the static solution
corresponding to the larger (right-most) zero of any such
pair is a local minimum of the energy -- a metastable vacuum --
while the 
solution corresponding to the left-most zero is a saddle-point, which
can be interpreted as an antikink perched at a distance from the
boundary at which the
force between it and the boundary vanishes,
unstable to decay in one direction to the metastable vacuum just
discussed, and in
the other to the next metastable
vacuum down ($n\to n{-}1$) with the antikink
escaping to minus infinity. (A similar phenomenon occurs in the
boundary $\phi^4$ theory with a suitably-signed boundary magnetic field
\cite{Dorey2015}.) As $k\to 0$ the perched antikinks corresponding to
the non-trivial parts of the
saddle-point solutions all move away to $x=-\infinity$, leaving only
the metastable vacua which become degenerate in energy
with the $u=0$ ground state as $k$ reaches $0$ and the Neumann
boundary is recovered.

\begin{figure}
\includegraphics[width = 0.94\linewidth]{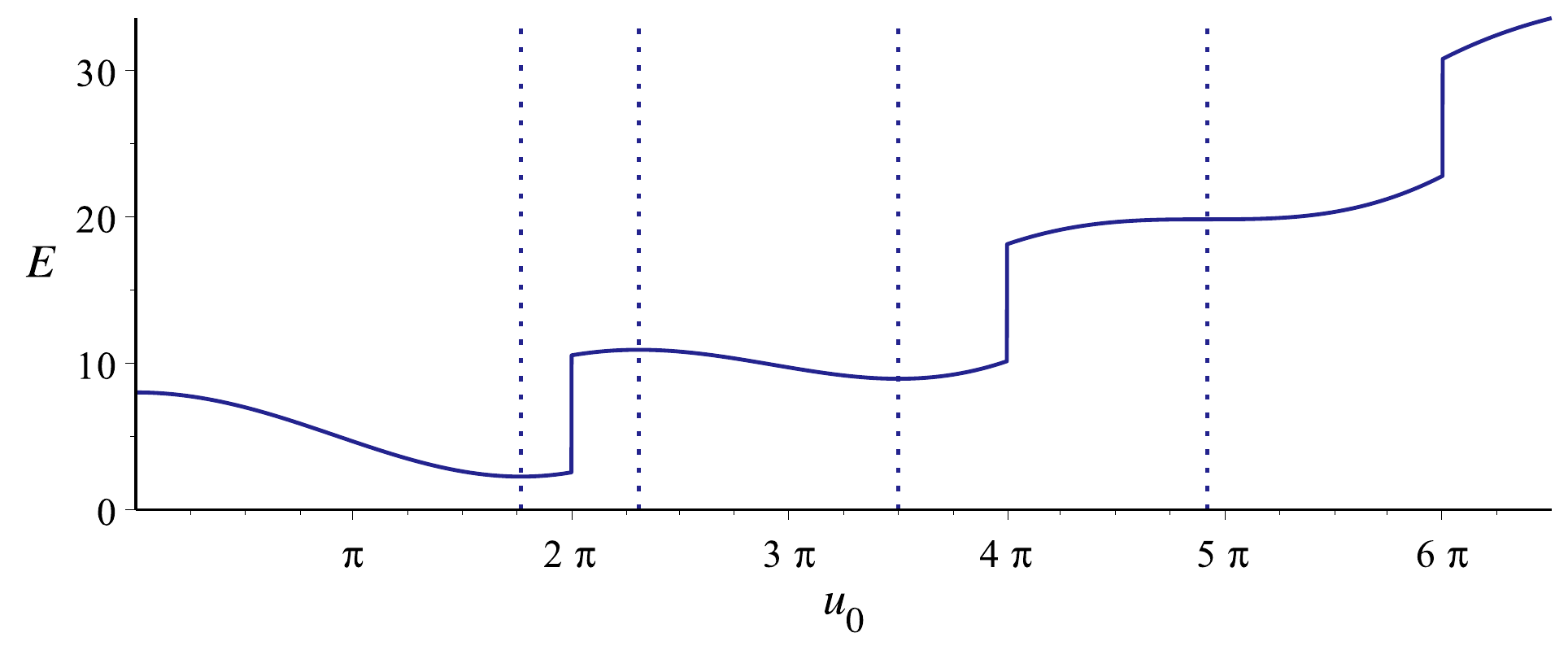}
\caption{\small 
The energy, $E$, of a static antikink $u(x)$ with $u(0)=u_0$ as given 
by \refeq{staticEnergy} with 
$k=0.064187$, the third critical value of $k$ shown in
\reffig{vacua}.
The vertical dotted lines indicate the solutions of \refeq{vaceq},
 which are also the stationary points of $E(u_0)$.
Note that if a point $u_0=2\pi m$ is approached from the left then
$u(x) \rightarrow 2\pi m$ everywhere, while the solution found when
approaching from the right contains a static antikink $u(x) =
4\arctan\left(\exp({x_0-x})\right) + 2\pi m$ whose position $x_0
\rightarrow -\infty$ in the limit.
The bulk energy contribution of $u(x)=2\pi m$ is zero while a static
antikink on the full line has energy $8$, so $E(u_0)$ has a
discontinuity of magnitude 8 every $u_0=2\pi m$.
}
	\label{vacuaEnergy}
\end{figure}

\subsection{Forces}
\label{Force}

\begin{figure}
	\centering
	\includegraphics[width=0.8\linewidth]{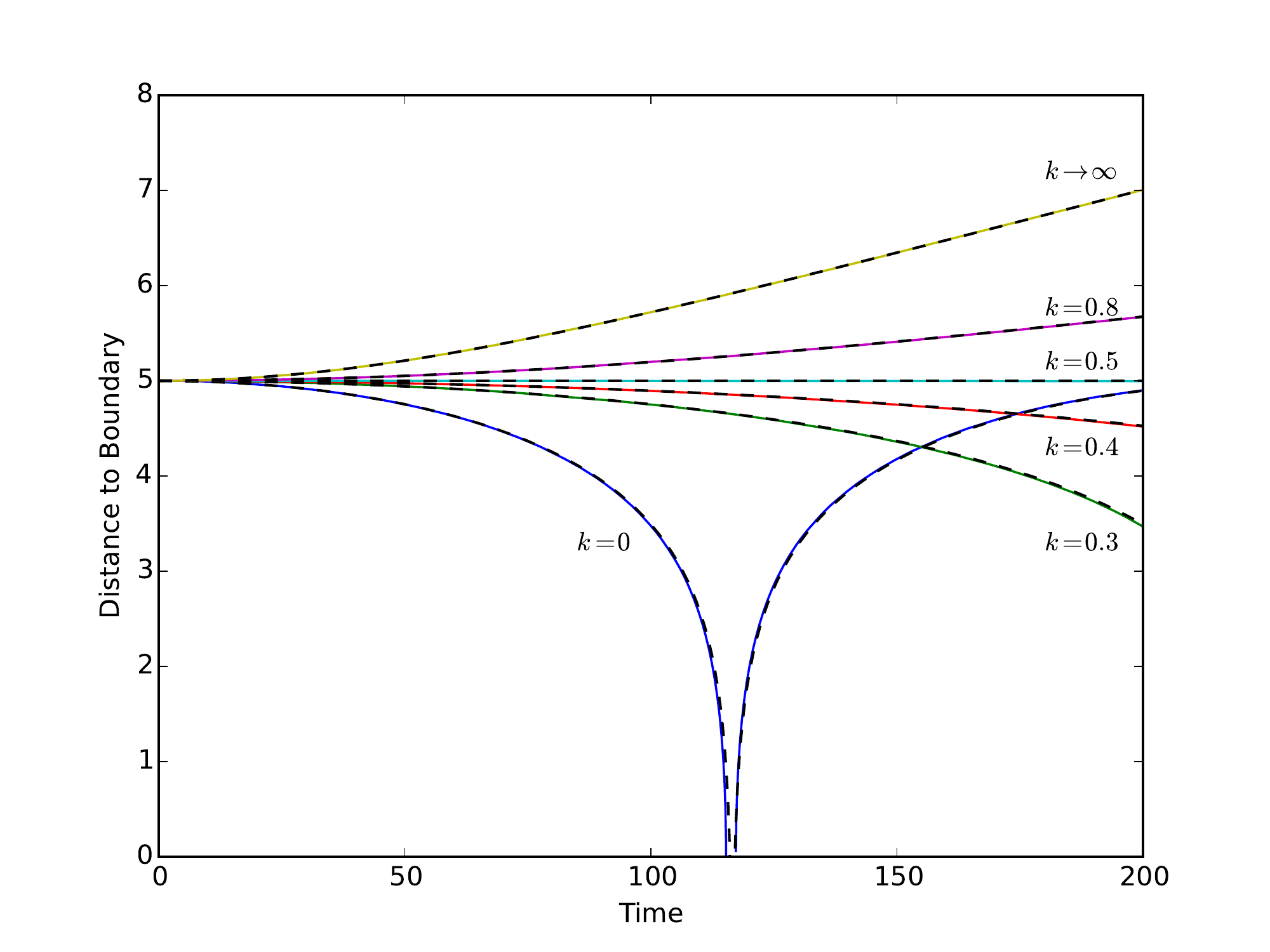}
	\caption{\small
Numerically-determined trajectories 
of an antikink with zero initial
velocity placed at $x=-5$ with a (ground state)
Robin boundary at $x=0$, with various
boundary parameter values $k$. After the collision of the $k=0$ trajectory the
position of the reflected kink is tracked instead. The dashed lines
show the trajectories as would follow from the force law
(\ref{robinforce}).}
	\label{antikinkTrajectory}
\end{figure}

We first consider solitons and breathers sitting to the left of a
Robin boundary in its 
ground state, so $u\approx 0$ in the vicinity of the boundary.
The asymptotic force on a static antikink at $x_0<0$ with $|x_0|\gg 1$
can be found as in \cite{Dorey2015}: we park
an `image' kink (or, for larger values of $k$, an antikink)
at $x_1>0$ in such a way that the
combined configuration satisfies
the Robin boundary condition at $x=0$, and then use the standard
full-line result that a sine-Gordon antikink and kink a 
distance $R\gg 1$ apart
experience an attractive force $F=32\,e^{-R}$ (see for example
\cite{Manton2004}). 
The antikink-kink combination can be approximated 
as
\begin{equation}
u(x) = 
4\arctan\left(e^{-(x-x_0)}\right)+ 4\arctan\left(e^{x-x_1}\right)
        \label{antikinkkink}
\end{equation}
so for $|x_0|$ and $|x_1|$ both large the Robin boundary condition
$u_x|_{x=0}+2ku|_{x=0}=0$ becomes
\begin{equation}
4(-e^{x_0}+e^{-x_1})+8k
(e^{x_0}+e^{-x_1})=0\,.
\end{equation}
Solving for $e^{-x_1}$ and computing the force yields
\begin{equation}
F=32\,e^{-(x_1-x_0)}=32\,\frac{1{-}2k}{1{+}2k}\,e^{2x_0}\,.
\label{robinforce}
\end{equation}
For $k>1/2$ an image antikink should be used instead, but the final
formula is unchanged, with the force now repulsive instead of
attractive.
In the integrable Neumann and Dirichlet
limits $k=0$ and $k\to\infty$ this result matches the asymptotic
behaviour of the corresponding exact solutions;  and
as shown in \reffig{antikinkTrajectory}, it also agrees well at
intermediate points, including the `critical' value $k_c=1/2$ at which
the predicted force vanishes.

\begin{figure}
	\centering
	\includegraphics[width=0.8\linewidth]{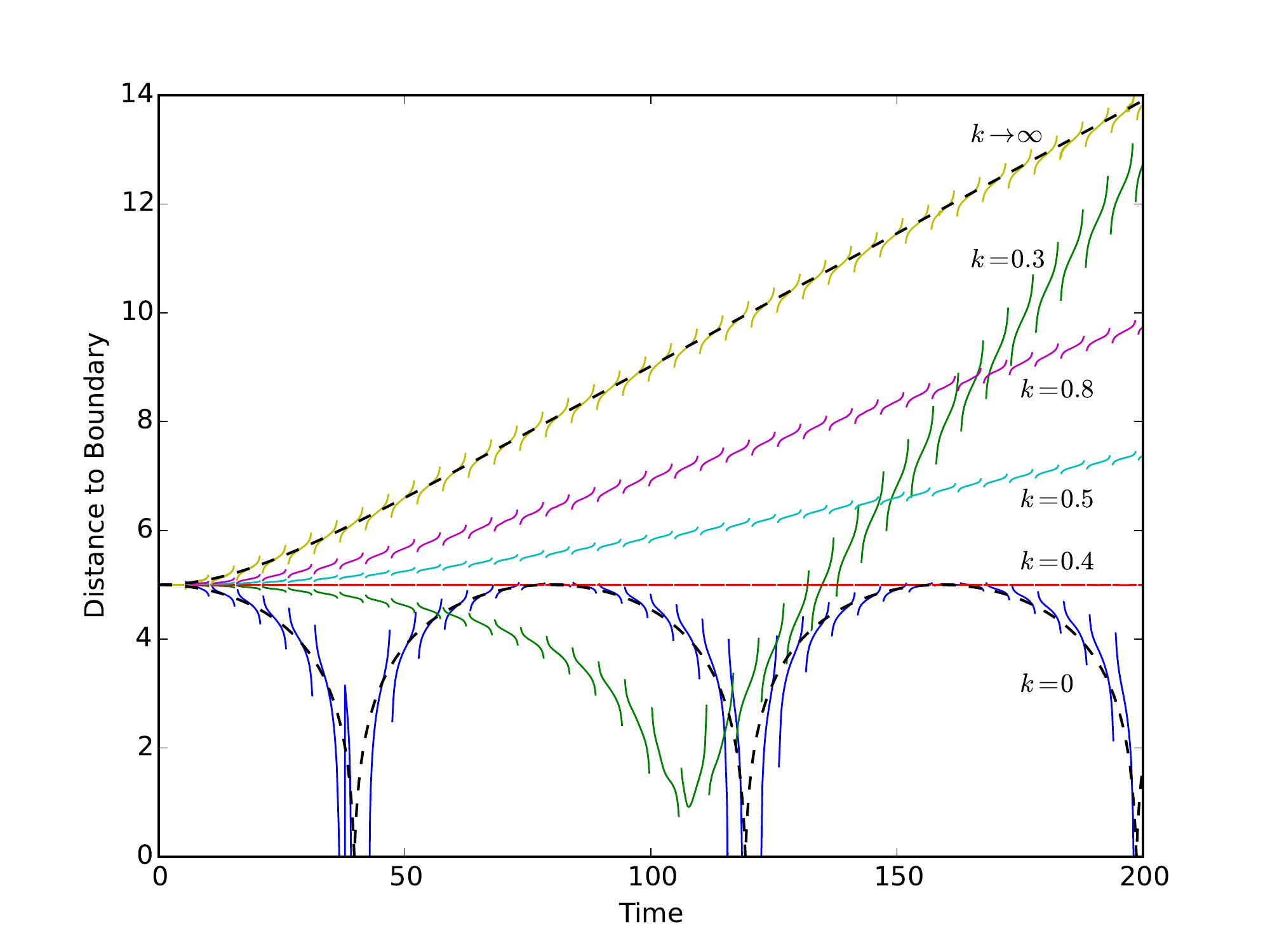}
	\caption{\small
Numerically-determined trajectories of a breather with zero initial
velocity and frequency 0.6
placed at $x = -5$ with a Robin boundary at $x=0$, for various values
of the boundary parameter $k$.
The position of the breather is 
defined as the point where the absolute value of the
field reaches its maximum, with points
where this is less than $1$ omitted for
clarity.  The dashed lines are the theoretical trajectories for
Dirichlet (top) and Neumann (bottom) boundaries.  These 
correspond to half the breather separation for two out of
phase and in phase breathers respectively, as calculated in
\cite{Nishida2009} using a collective coordinates method.}
	\label{breatherTrajectory}
\end{figure}

The situation is more subtle for a breather placed near the
Robin boundary, but at least the limiting integrable cases are
straightforward: they can be modelled on the full line
by adding a symmetrically-placed image breather, 
exactly in phase with the `real'
breather for the Neumann boundary, and exactly out of phase for
Dirichlet.  Since it
can be shown that two in phase breathers feel an attractive
force while two out of phase breathers experience a repulsive force
\cite{Nishida2009} (results which we verified by
constructing the relevant exact two-breather solutions, as in
\cite{Dmitriev2001}), a stationary breather
is attracted by the $k=0$ boundary, while for $k=\infty$ it is repelled.
We do not have an analytical result for the general Robin
boundary, but we found numerically that
a similar interpolating behaviour emerges
as for the stationary kink or antikink, as shown in 
\reffig{breatherTrajectory}. 

We therefore conclude that the Neumann boundary is repulsive and the
Dirichlet attractive both for kinks and antikinks
 and for breathers, with the
homogeneous Robin 
boundary based on the $u=0$ vacuum transitioning smoothly from
attractive to repulsive as $k$ increases from 0 to infinity.  
However the critical value of $k$ at which the force vanishes is
different in the two cases: for breathers our numerical results show
that it
is frequency-dependent, only tending (from below) to the kink and
antikink  value  
$k_c=1/2$ as the frequency tends to zero.
It would be interesting to analyse this 
asymptotic breather-wall force 
in more detail, 
but we will leave this question for future work.

\begin{figure}
\centering
\includegraphics[width=0.8\linewidth]{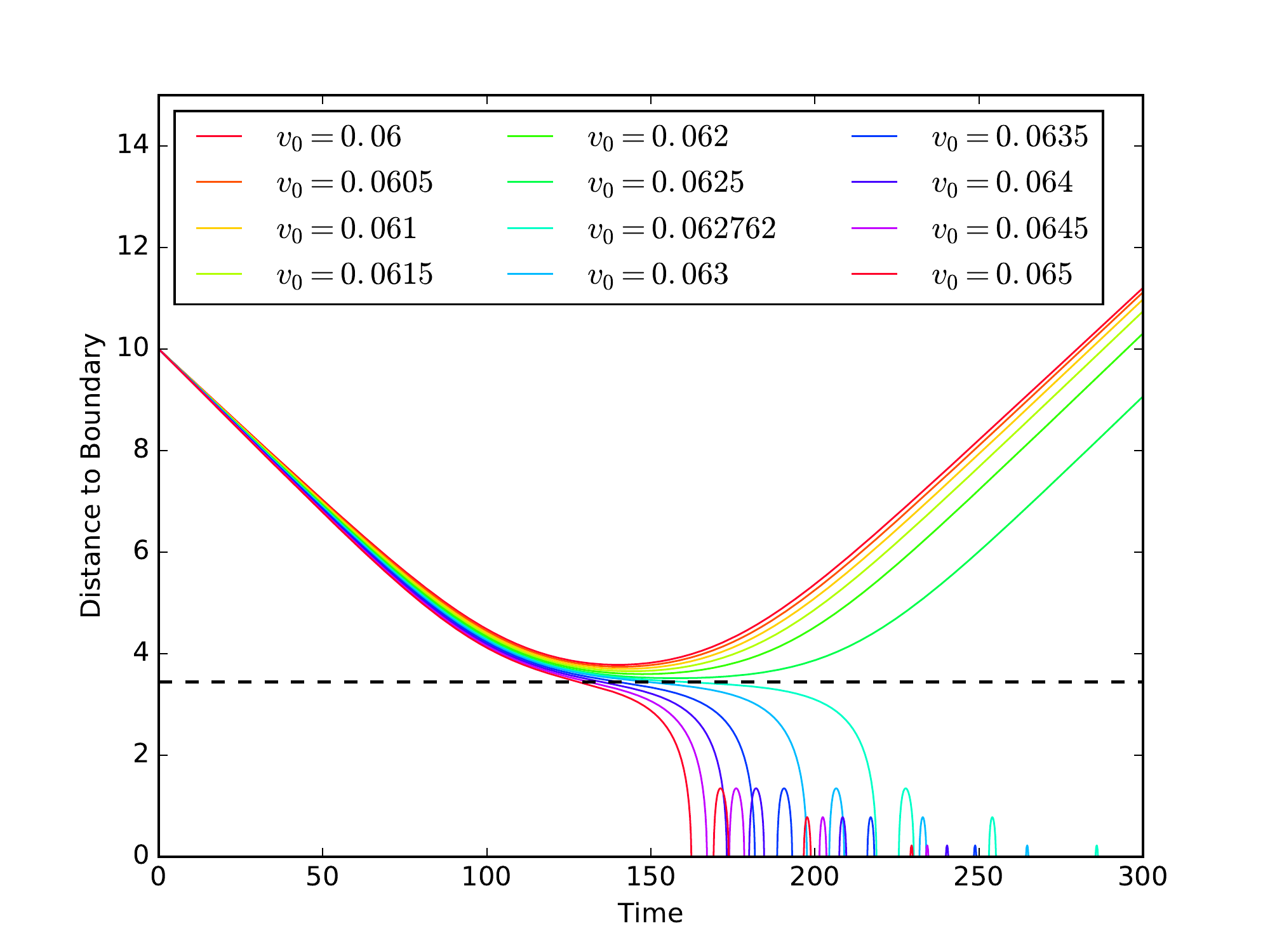}
\caption{\small
Numerically-determined 
trajectories of an antikink with various initial
velocities incident from $x_0=-10$ on the $n=1$ metastable boundary
at $x=0$, with $k=0.01$. The horizontal dashed line shows the distance
at which the force vanishes. An approximation based on comparing the
energy of a distant antikink with velocity $v_0$ with that of a
static antikink at the zero-force point predicts 
the transition from reflection to capture should occur at 
$v_0=0.062762$, in good
agreement with the numerical results.
}
        \label{2piantikinkTrajectory}
\end{figure}

For kinks, antikinks and breathers built on one of the metastable
vacua discussed in \refsec{Vac}, the story is considerably
more involved and we only have preliminary results. As in \refsec{Vac}
these vacua can be labelled by an integer $n$, so that $2\pi n$ is
the value that the field takes as $x\to -\infty$ in the 
absence of any additional kinks or antikinks. Supposing for ease
of exposition that $n$ is positive,
this vacuum
configuration can be modelled on the full line
by placing a single `image' antikink 
at some location $x_1>0$.
If a real kink or antikink is added at some $x_0<0$
(so that the limiting field value as $x\to
-\infty$ is now $2\pi(n{\pm}1)$) then so long as $x_0$ is sufficiently
negative, the combined
full-line kink-antikink or 
antikink-antikink configuration will continue to satisfy the
boundary condition with only a small change in $x_1$.
Hence a distant antikink will be repelled by a
metastable boundary with $n>0$, and a kink will be attracted.
However the situation changes for
the antikink when it gets closer to the boundary: the position at which
the image antikink must lie in order for the boundary condition to be
satisfied grows, diverging to infinity at the moment when the
real antikink on its own satisfies the boundary condition and hence
experiences no force, replicating
the unstable saddle-point solution that tends to $2(n{+}1)\pi$ as $x\to
-\infty$. At nearer distances still, the antikink is attracted towards
the boundary. This scenario is illustrated for the $n=1$ metastable
vacuum in
\reffig{2piantikinkTrajectory}. The horizontal dashed line shows
the zero-force distance $-x_0=3.439\dots$ from the
boundary, where $x_0=\ln(\tan(u_0/4-\pi/2))$ is the antikink location
in the relevant unstable static solution, with $u_0$ the solution to
\refeq{vaceq} in the interval $[2\pi,3\pi]$ for $k=0.01$.

For breathers the situation is, perhaps not surprisingly, even more
complicated. However our numerical simulations for the $n=1$
metastable vacuum show that while $k$ remains less than about $0.3$ 
and for breather frequencies around $0.6$ (typical for 
intermediate breathers in the processes we will discuss below) the
force is always attractive, confirming the apparent behaviour of intermediate
breathers in the spacetime plots of Figs.\,\ref{spaceTimeMess}a--g and 
\ref{spaceTimePhase}c below.

\section{Results for $k>0$}
\label{RobinResults}

\begin{figure}
	\includegraphics[width = 0.85\linewidth]{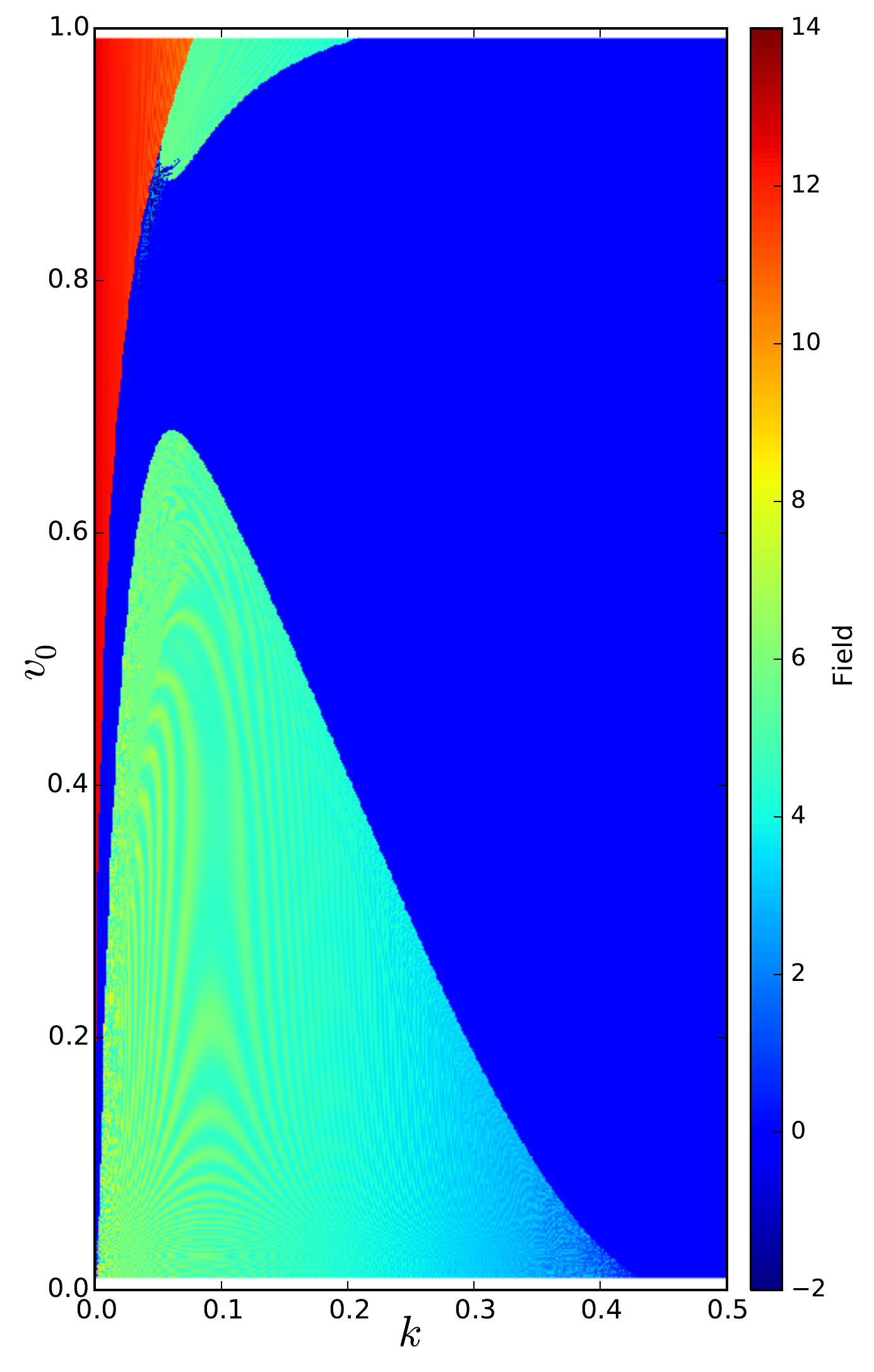}
	\caption{\small
 A snapshot of the field values at $x=0$, $t=|x_0|/v_0+1000$ for the
scattering of an initial antikink with velocity $v_0$, position $x_0$
and boundary parameter $k$. 
\reffig{snapshotMess} below shows a zoomed-in view of the 
complicated structure near to $k=0.06$, $v_0=0.89$.
	}
	\label{snapshotFull}
\end{figure}

\begin{figure}
\centering
\begin{subfigure}{0.8\textwidth}
  \centering
  \includegraphics[width=1\linewidth]{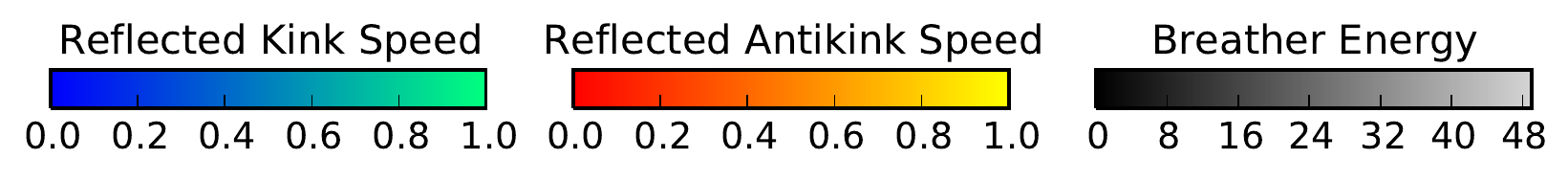}
\end{subfigure}
\begin{subfigure}[t]{0.485\textwidth}
  \centering
  \hspace*{-15pt}
  \includegraphics[width=1.07\linewidth]{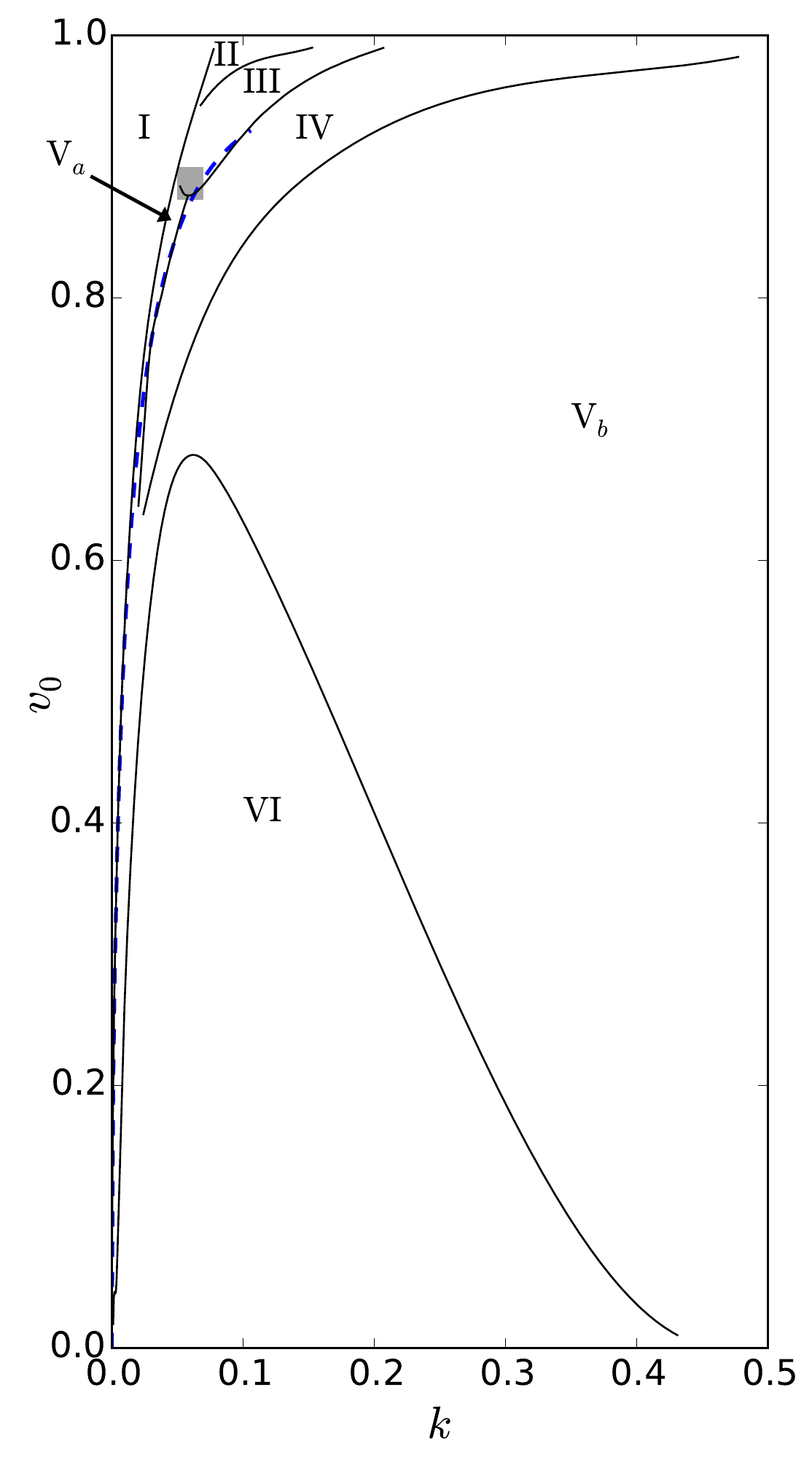}
  \caption{\small
Final states classified by kink, antikink and high energy
breather content: \\[2pt]
I: Kink\\
II: Kink and antikink\\
III: High-energy breather\\
IV: High-energy breather and antikink\\
V$_a$ \& V$_b$: Antikink\\
VI: None of the above.
}
  \label{map}
\end{subfigure}
\hfill
\begin{subfigure}[t]{0.485\textwidth}
  \centering
  \hspace*{-15pt}
  \includegraphics[width=1.07\linewidth]{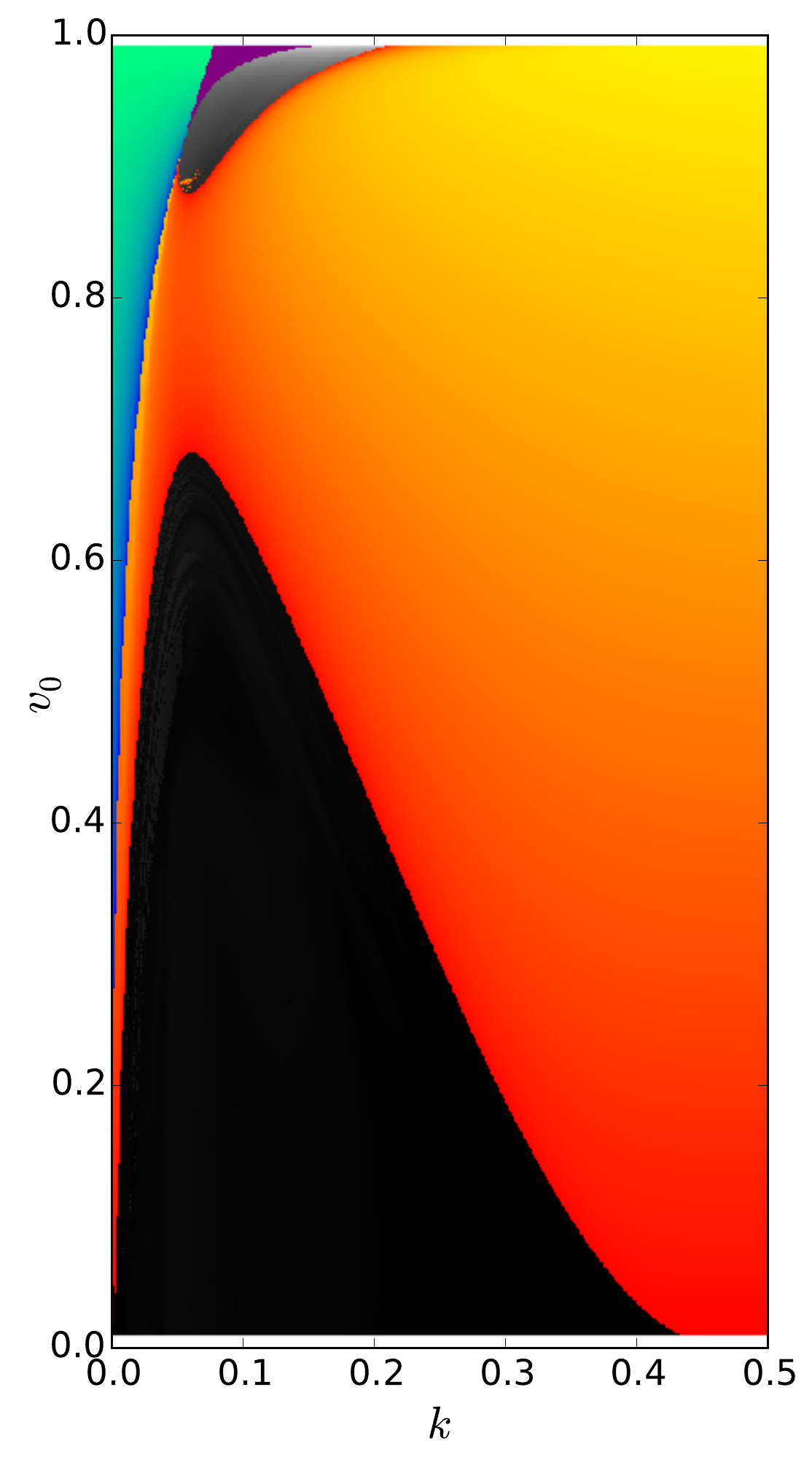}
  \caption{\small
Final state kinematics: If the final state contains a single
kink or antikink, its
speed is plotted; if neither, then the
total energy of all breathers detected in the final state is shown instead.
In the solid purple region the final state contains a kink and an
antikink.  
}
  \label{colourMap}
\end{subfigure}
\caption{\small
Possible outcomes of the collision of an antikink with initial speed $v_0$
with a Robin boundary with parameter $k$.  
In nonintegrable cases there is also some radiation in the final state.
The small shaded region 
in \reffig{map} 
is scrutinised in more detail in
\reffig{snapshotMess} and \refsec{BreatherRebound} below;
the dashed line shows the outer
limit for region I which is
derived in \refeq{energybound}. }
\label{maps}
\end{figure}

\begin{figure}
\begin{subfigure}{0.99\textwidth}
  {\small a) $k=0.05$,~ $v_0=0.95$ ~(region I)}\\[-4pt]
  \includegraphics[width=0.48\linewidth]{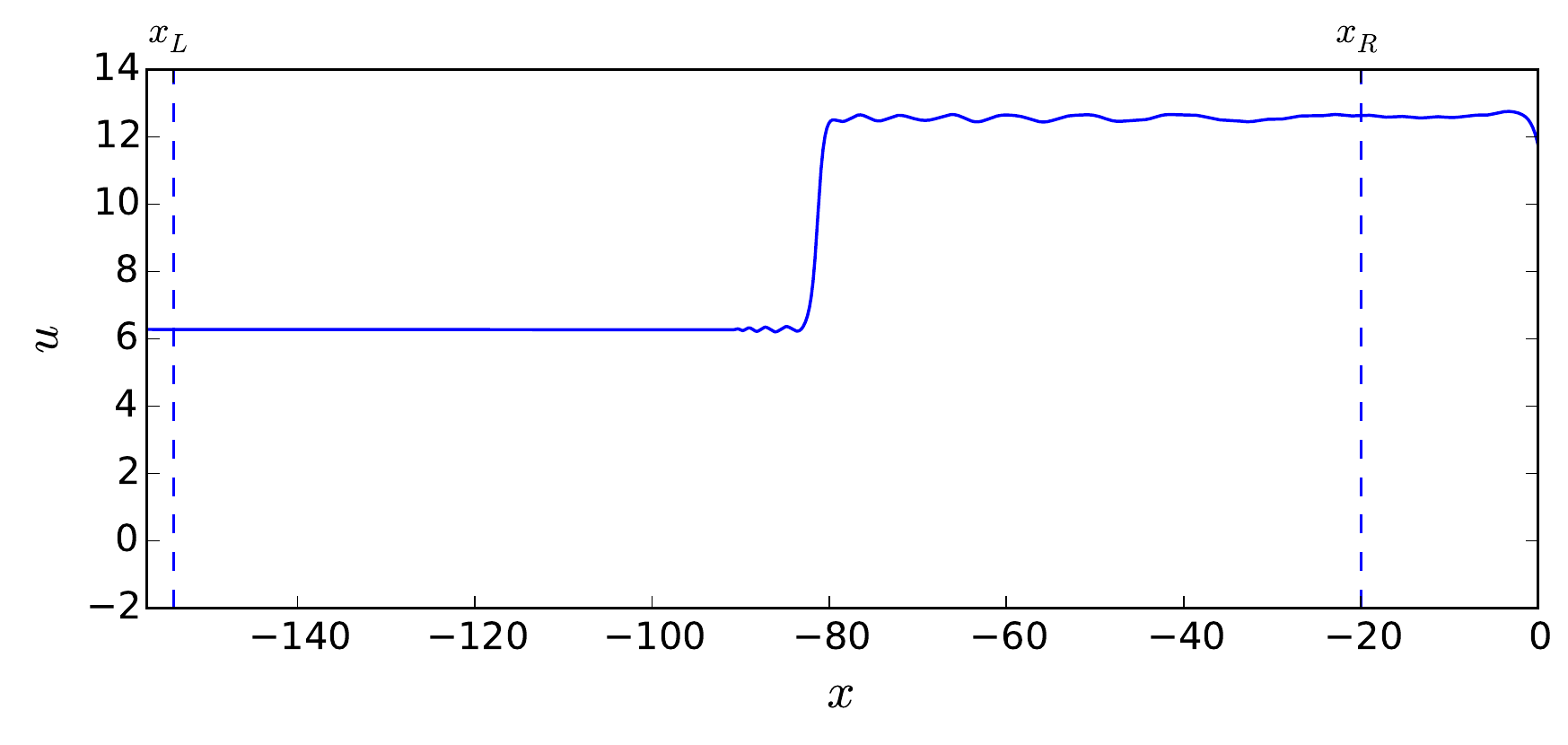}~~~
  \includegraphics[width=0.46\linewidth]{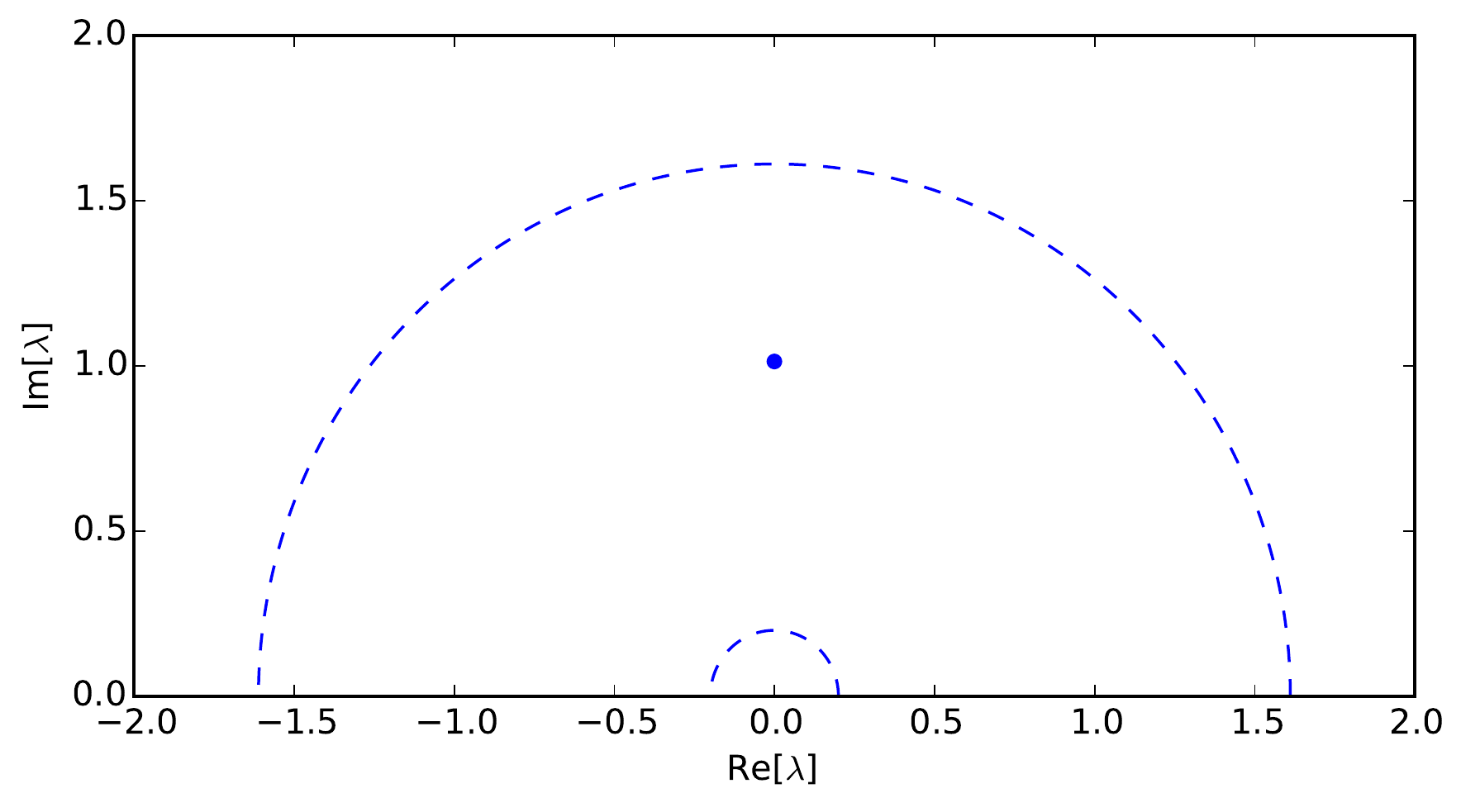}
  \label{k0p05}
\end{subfigure}
\\[10pt]
\begin{subfigure}{0.99\textwidth}
  {\small b) $k=0.065$,~ $v_0=0.95$ ~(region II)}\\[-4pt]
  \includegraphics[width=0.48\linewidth]{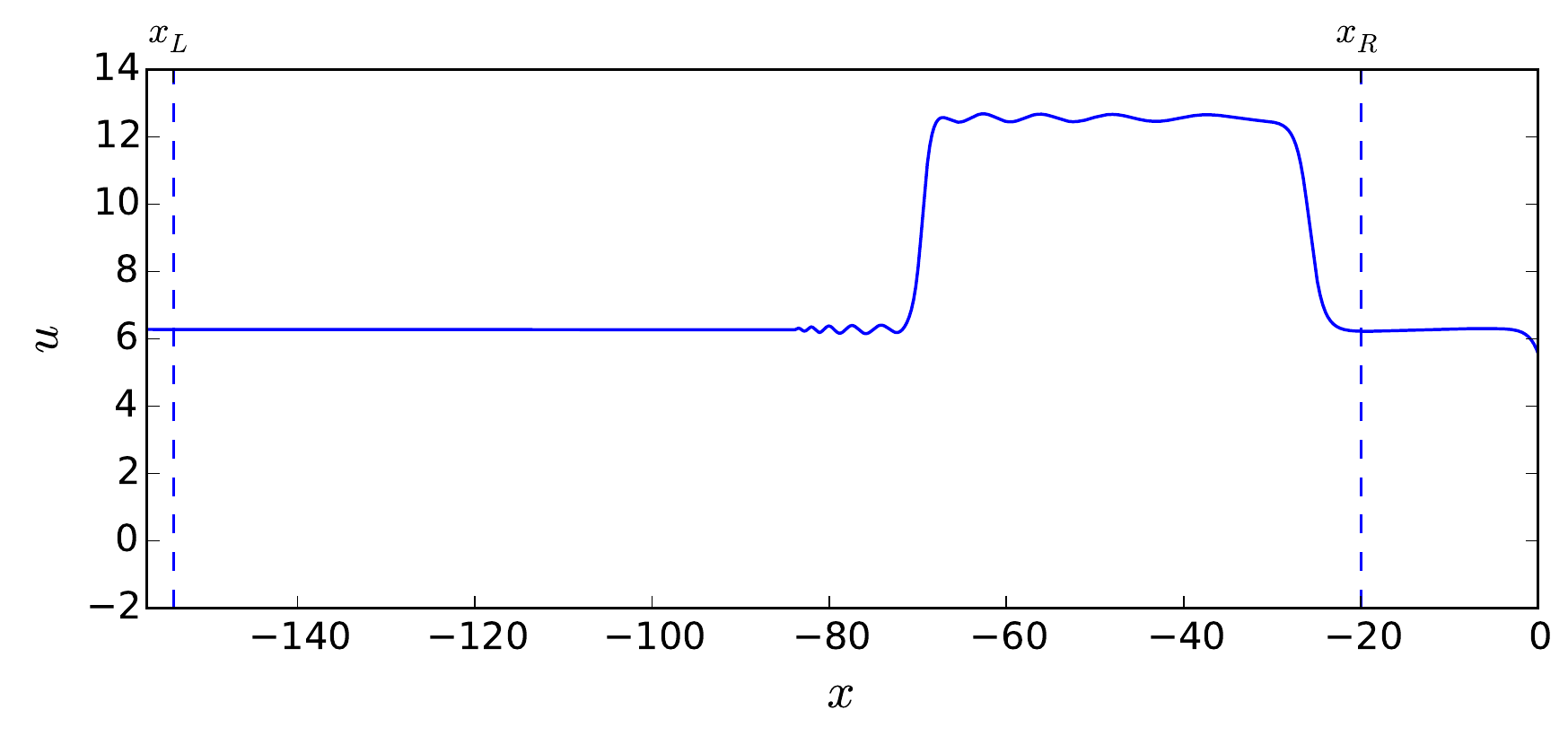}~~~
  \includegraphics[width=0.46\linewidth]{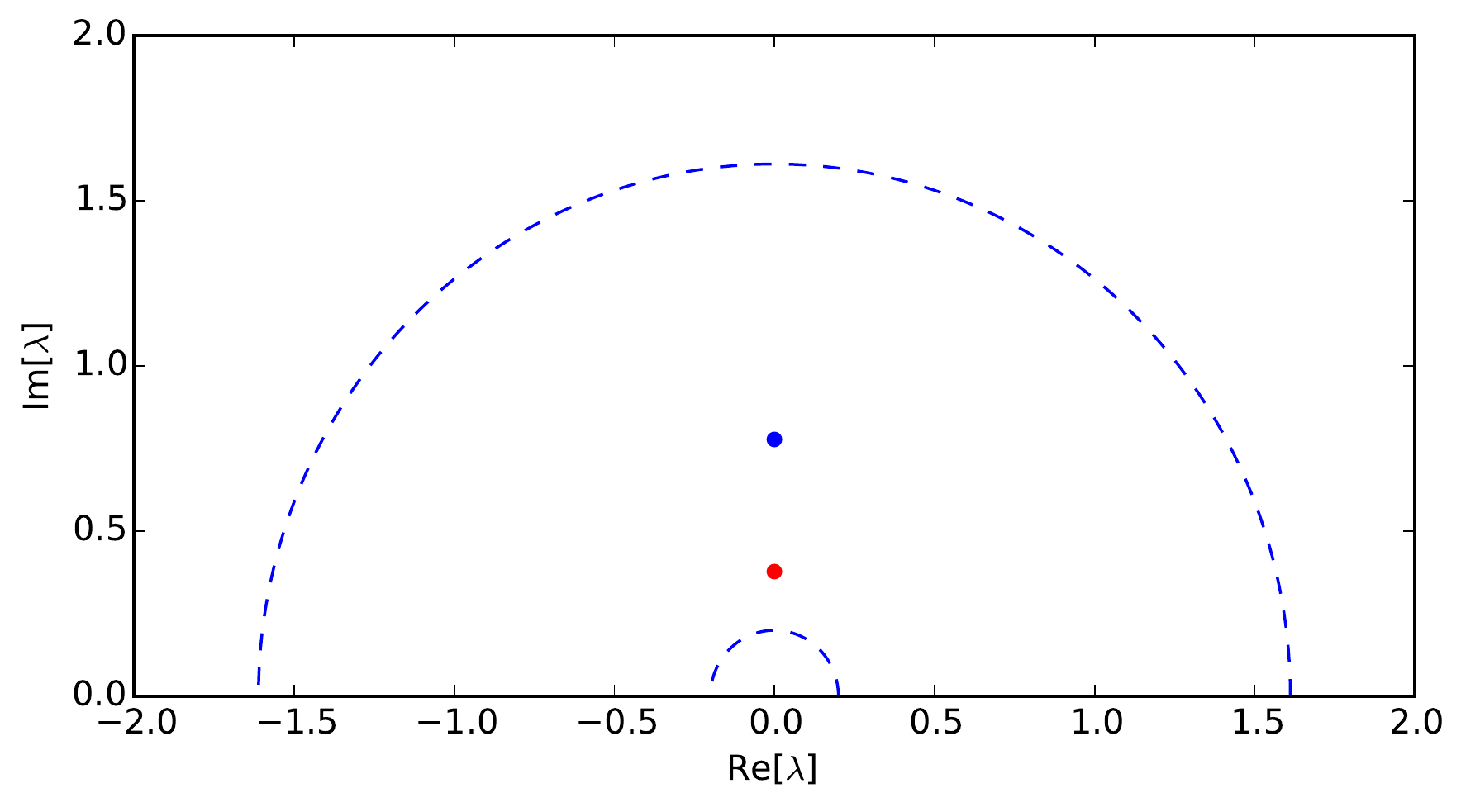}
  \label{k0p065}
\end{subfigure}
\\[10pt]
\begin{subfigure}{0.99\textwidth}
  {\small c) $k=0.09$,~ $v_0=0.95$ ~(region III)}\\[-4pt]
  \includegraphics[width=0.48\linewidth]{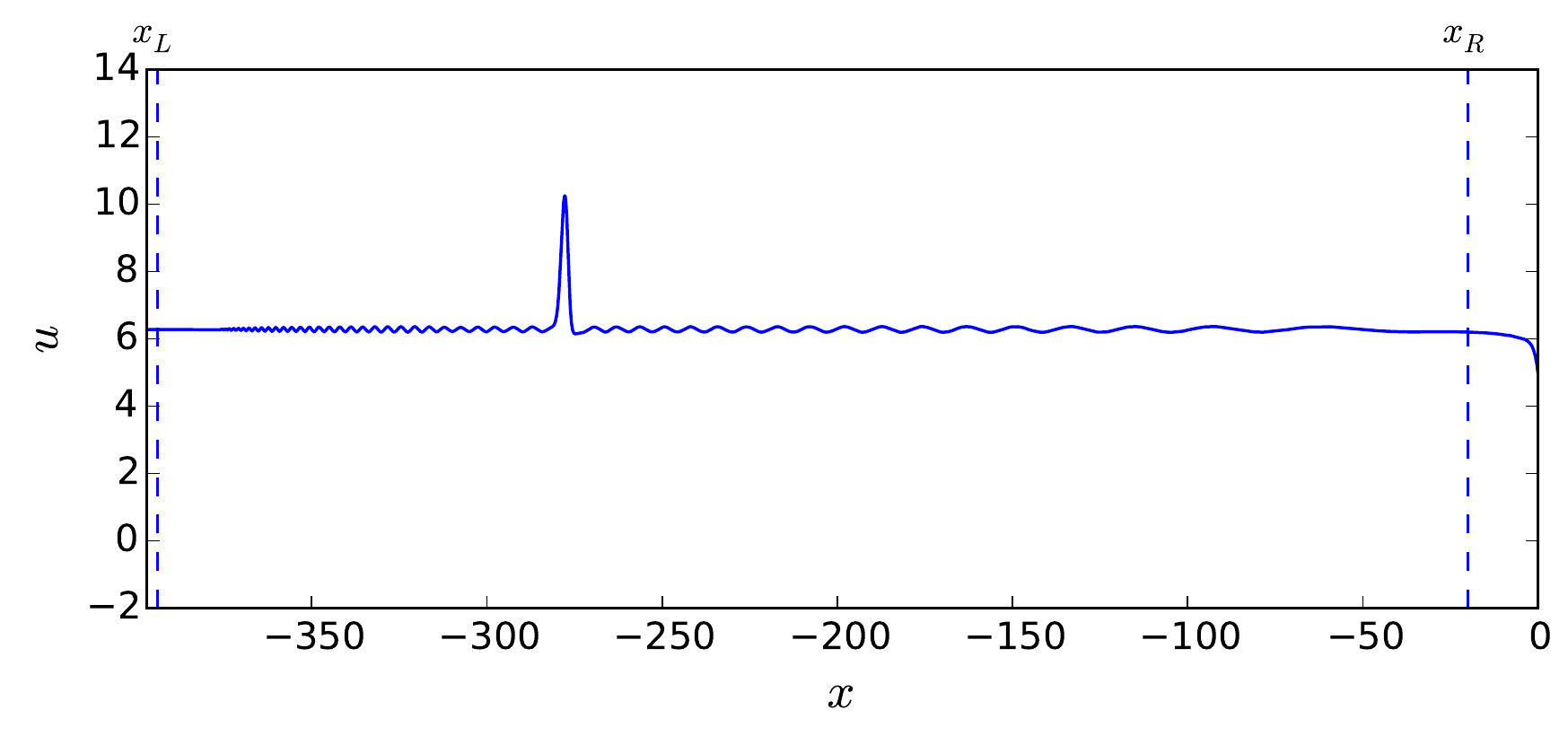}~~~
  \includegraphics[width=0.46\linewidth]{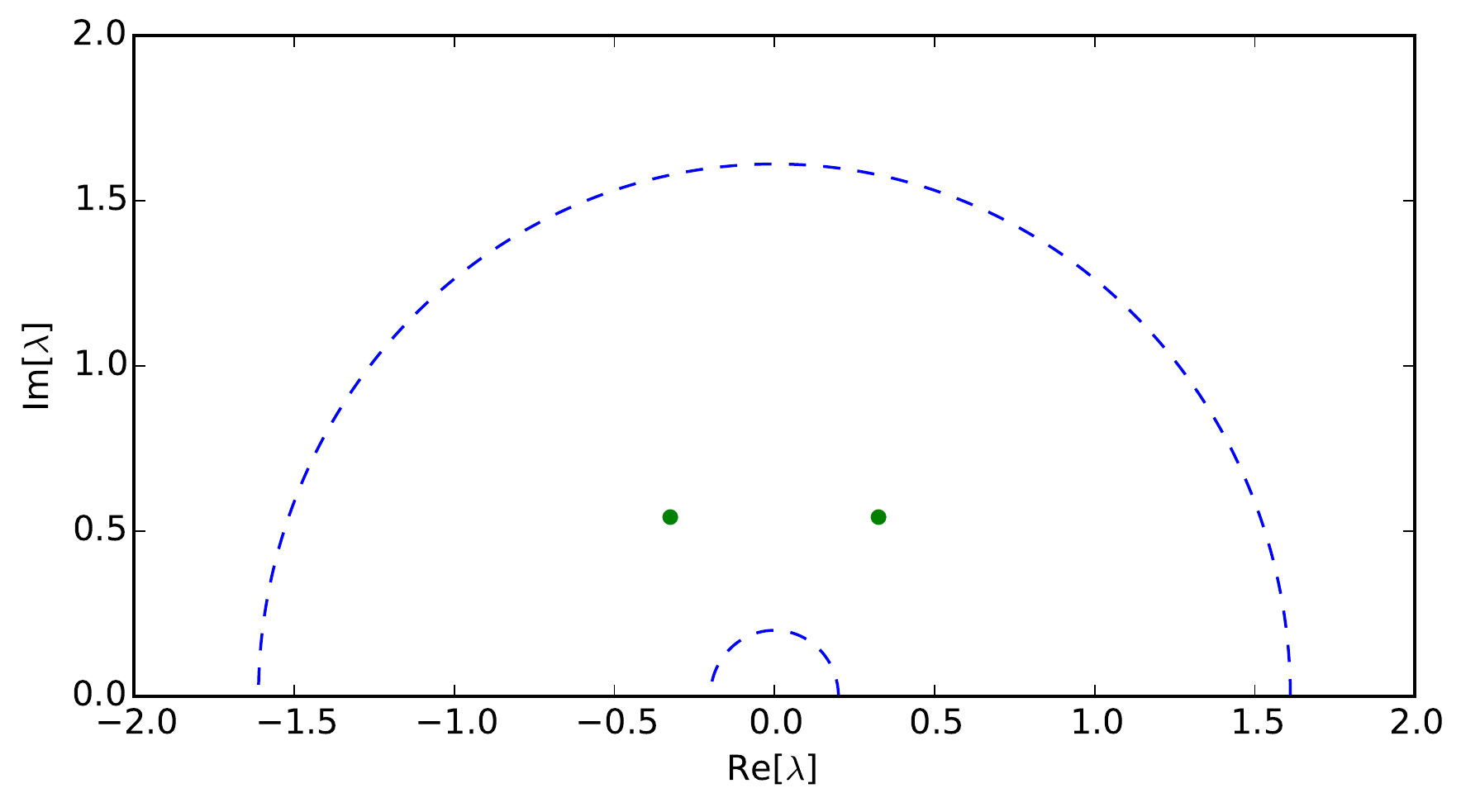}
  \label{k0p09}
\end{subfigure}
\\[10pt]
\begin{subfigure}{0.99\textwidth}
  {\small d) $k=0.145$,~ $v_0=0.95$ ~(region IV)}\\[-4pt]
  \includegraphics[width=0.48\linewidth]{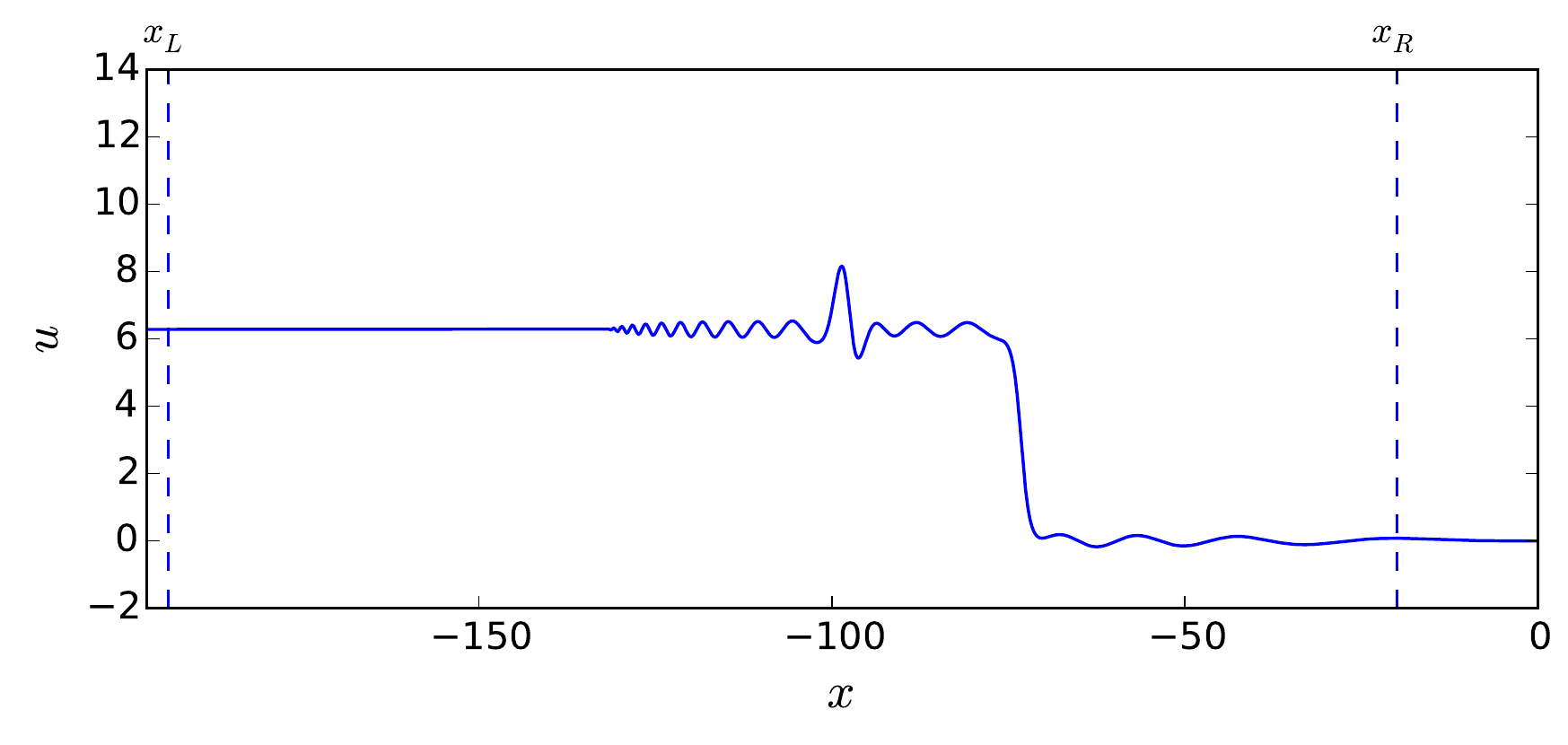}~~~
  \includegraphics[width=0.46\linewidth]{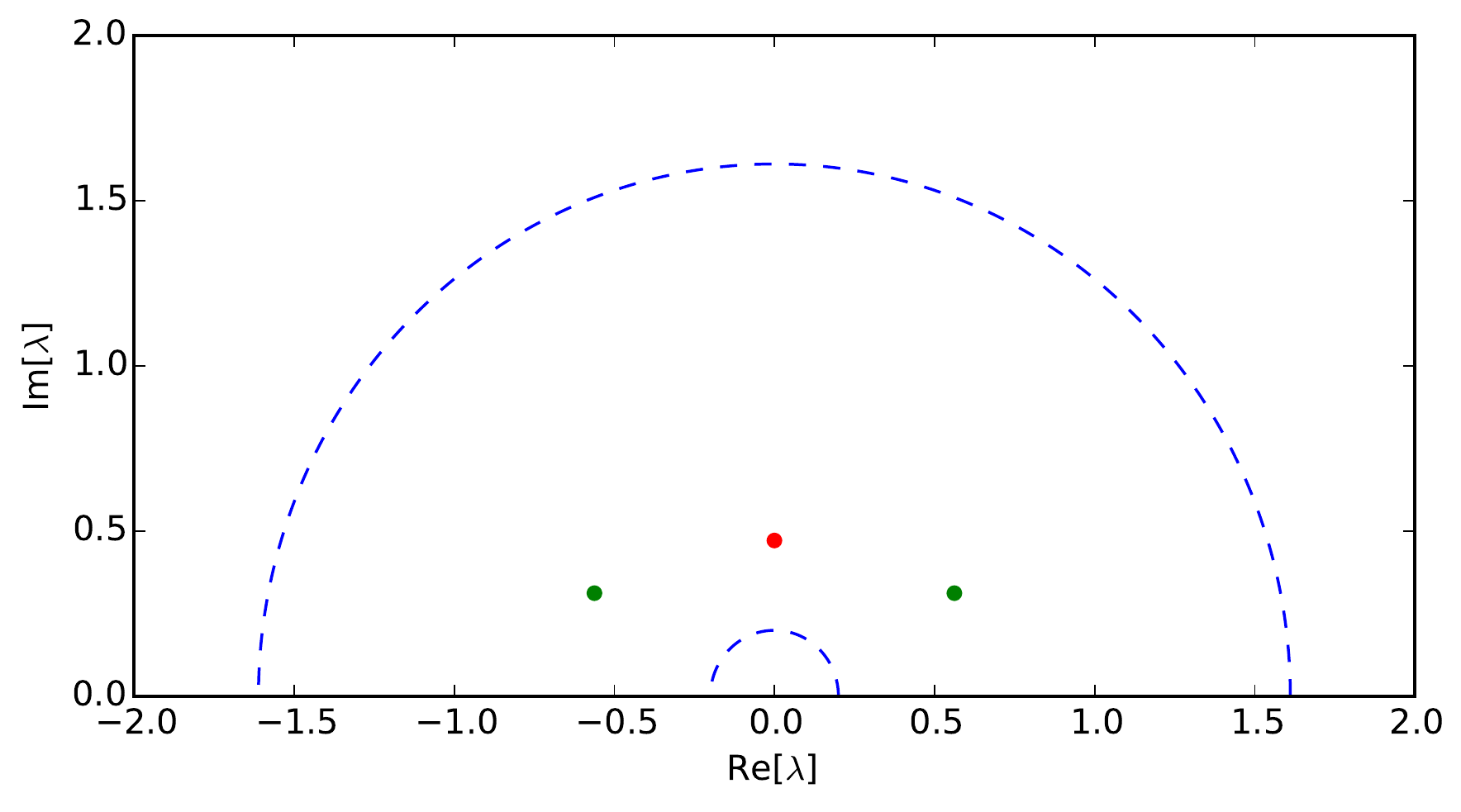}
  \label{k0p145}
\end{subfigure}
\\[10pt]
\begin{subfigure}{0.99\textwidth}
  {\small e) $k=0.3$,~ $v_0=0.95$ ~(region V$_b$)}\\[-4pt]
  \includegraphics[width=0.48\linewidth]{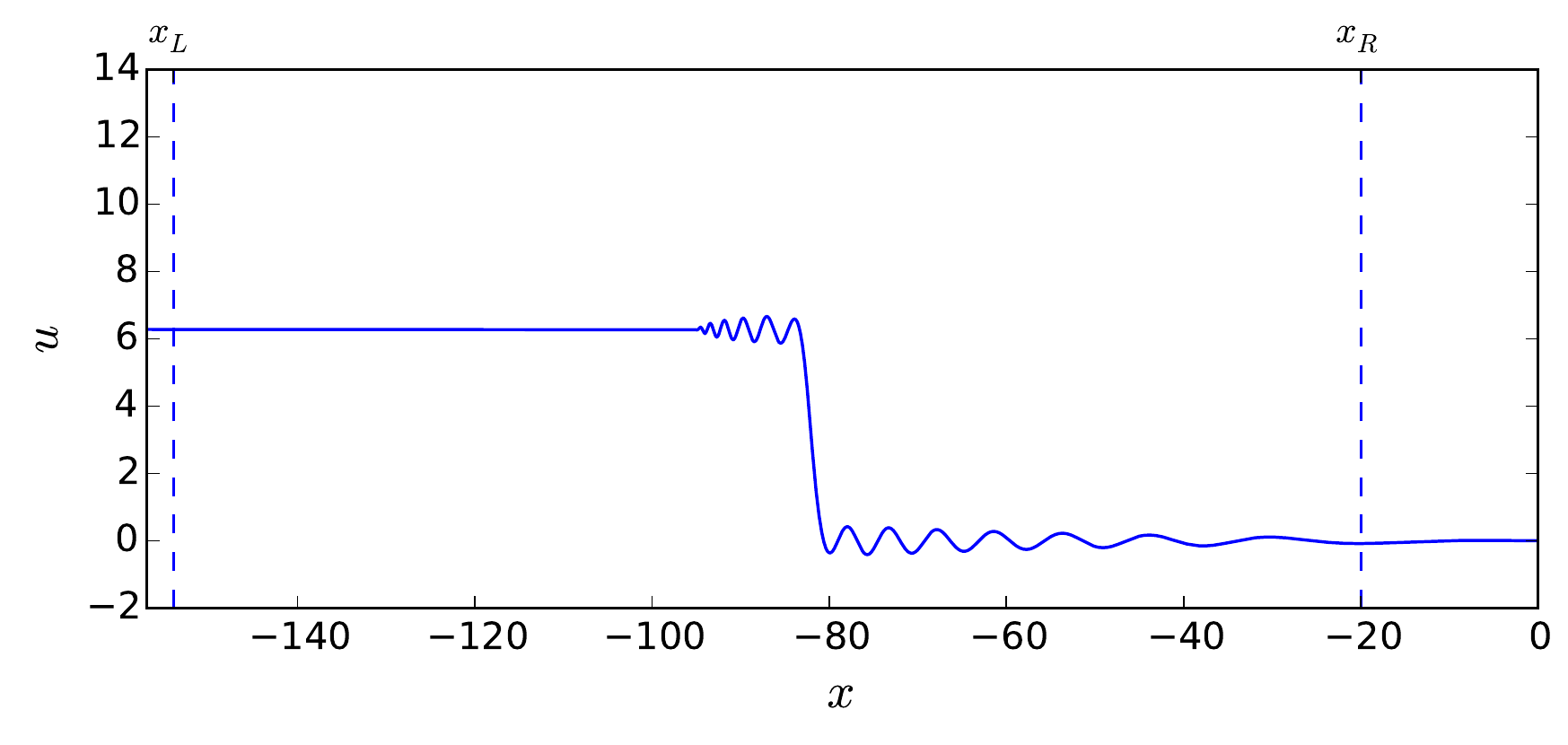}~~~
  \includegraphics[width=0.46\linewidth]{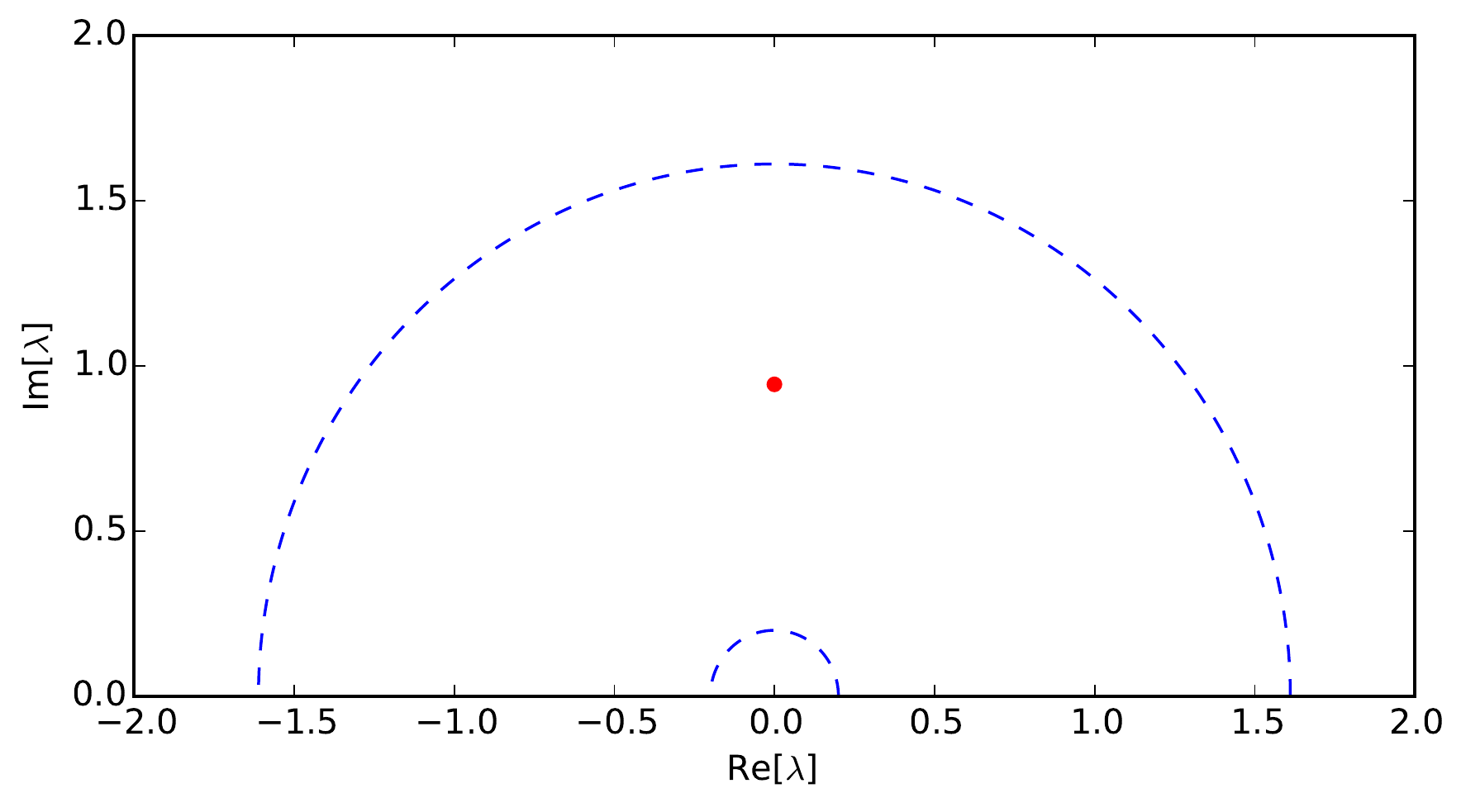}
  \label{k0p3}
\end{subfigure}
\caption{\small
Scattered field (left) and bound state eigenvalues (right) for a
sequence of values of $k$, all with $v_0=0.95$, illustrating how the
eigenvalues evolve with changing $k$.
}
\label{profilesandeigenvalues}
\end{figure}

\begin{figure}
	\includegraphics[width=\linewidth]{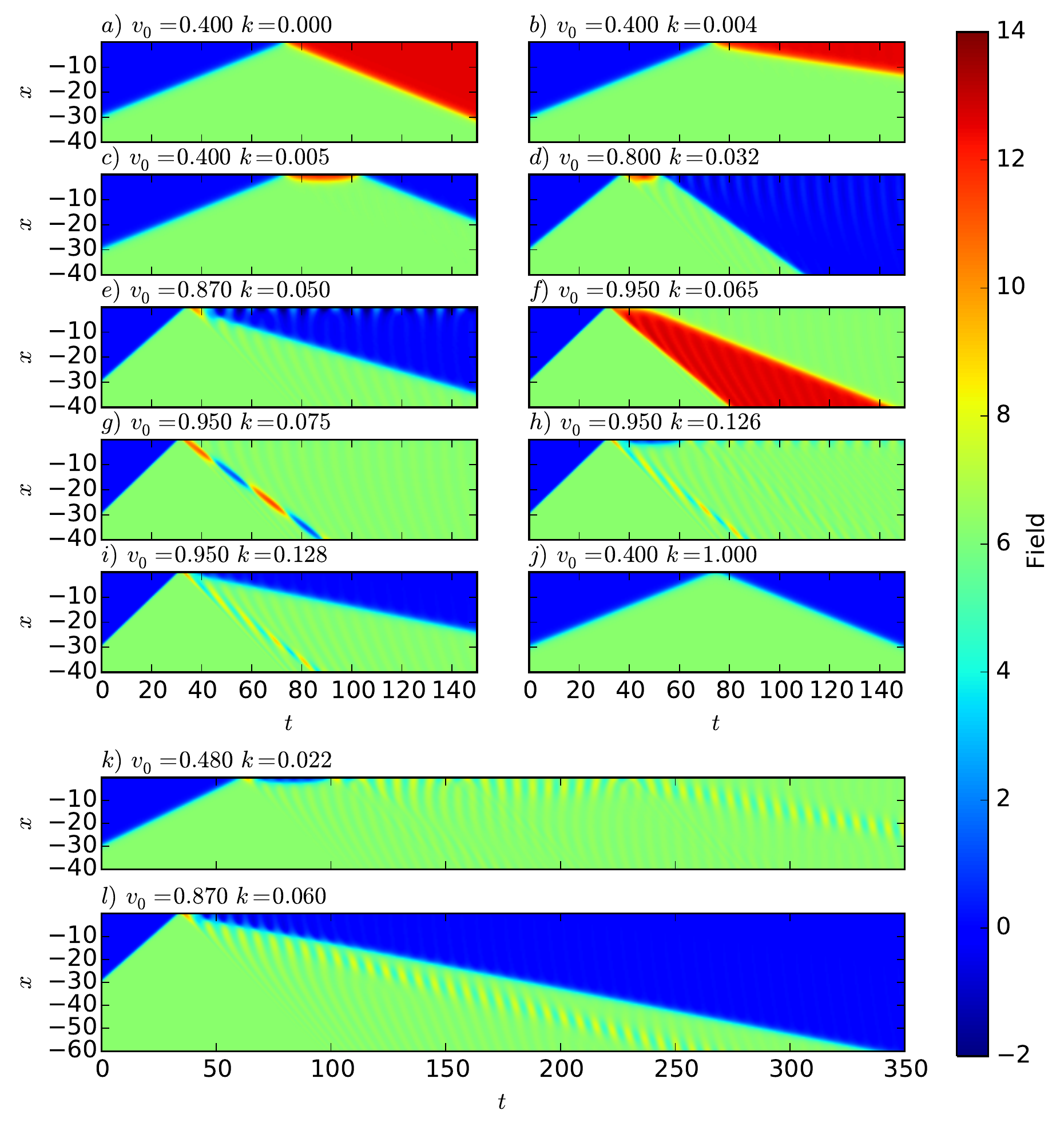}
	\caption{\small
Spacetime plots showing the collision of an antikink with initial
velocity $v_0$ with the Robin boundary (\ref{robin}).
        The types, velocities $v$ and frequencies
$\omega$ of the excitations produced by the collisions, excluding breathers with $\omega > 0.999$,  are:
        a) a kink with $v=-0.400$;
        b) a kink with $v=-0.149$;
        c) an antikink with $v=-0.391$;
        d) an antikink with $v=-0.69$ and breather with $v=-0.107$, $\omega=0.996$;
        e) an antikink with $v=-0.29$;
        f) an antikink with $v=-0.40$ and a kink with $v=-0.81$;
        g) a breather with $v=-0.710$, $\omega=0.30$;
        h) a breather with $v=-0.72$, $\omega=0.78$;
        i) an antikink with $v=-0.2$ and breather with $v=-0.722$ and $\omega=0.80$;
        j) an antikink with $v=-0.400$;
        k) a breather with $v\approx-0.1$, $\omega\approx 0.93$;
        l) an antikink with $v=-0.195$ and breather with $v=-0.26$, $\omega=0.93$.
The numbers of digits quoted give a rough estimate of the accuracy of
the results for each plot, based on the 
extent to which they had stabilised by the time
the finest grid of $dx=0.0025$, $dt=0.002$ was reached. 
}
	\label{spaceTime}
\end{figure}

We now return to the collision of a single antikink with a
$k>0$ homogeneous
Robin boundary, and the way that this process interpolates
 between
the integrable $k=0$ and $k\rightarrow\infinity$ limits.
Many features of these collisions can be deduced from
\reffig{snapshotFull},
a simple `snapshot' plot
of the field values at the boundary a fixed time
after the initial impact.  Two further plots in
\reffig{maps} summarise the results of a more-detailed analysis based
on the numerical solution of the direct scattering problem for the 
final-state field, classifying the final states by their kink,
antikink and high-energy breather content. Away from the integrable
limits the final state always contains some radiation; and
in various areas of the
phase diagram (for example in parts of region VI) we also detected
numbers of low-energy breathers. However the pattern of these
low-energy breathers seems to be rather intricate, and hard to
distinguish numerically from radiation, as the corresponding zeros of the
Wronskian $W(\lambda)$ lie very near to the real axis. For this
reason we will not discuss the low-energy breathers
in detail below, but it would be of
interest to return to their study in the future.
Some typical examples of final states and the corresponding patterns
of Wronskian zeros are shown in
\reffig{profilesandeigenvalues}. Finally
\reffig{spaceTime} illustrates some of the processes involved 
via spacetime plots of a variety of special cases.

\begin{figure}
	\includegraphics[width=\linewidth]{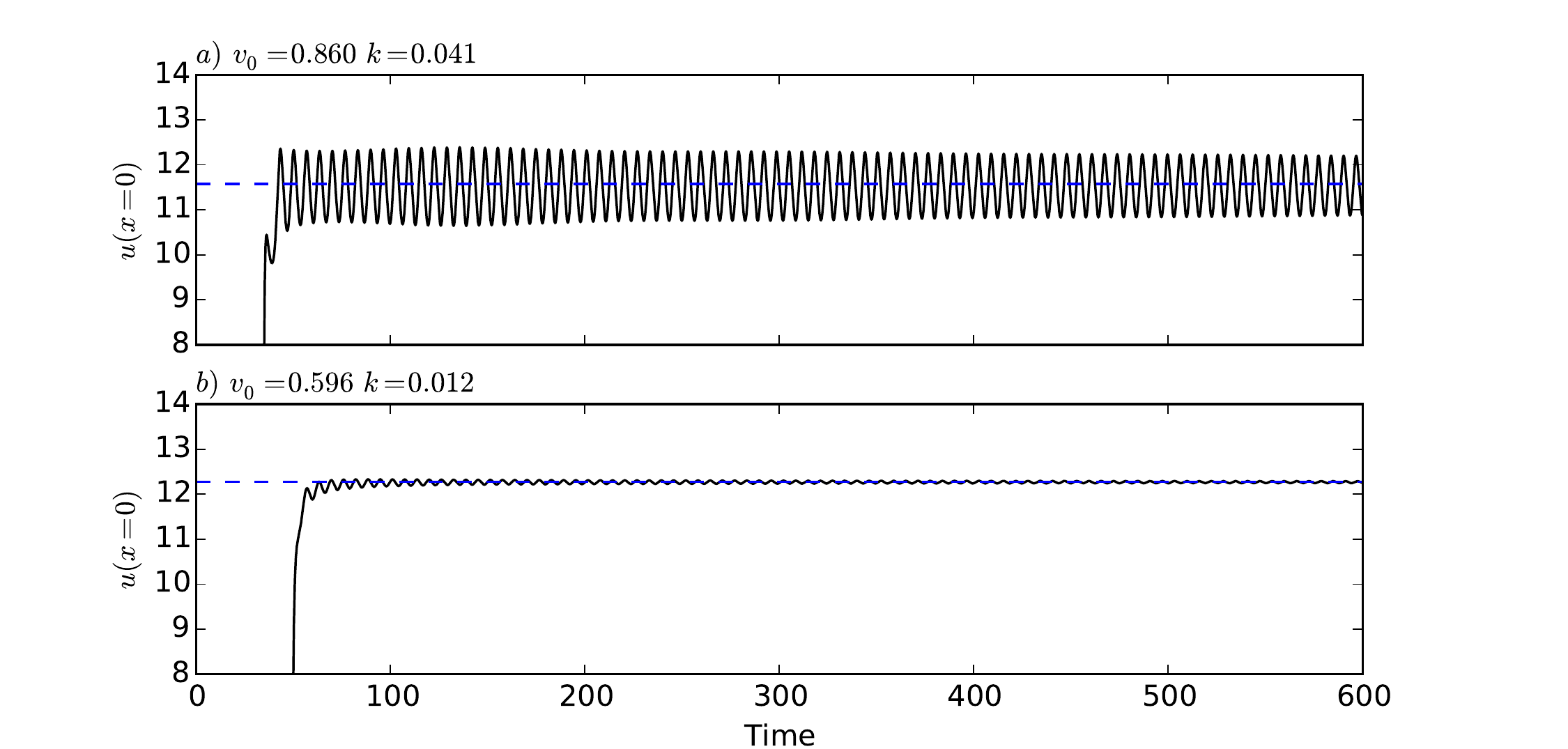}
	\caption{\small
The solid line is the value of the field at the boundary, $u(x{=}0,t)$, 
as an antikink with initial velocity $v_0$ collides with the Robin
boundary with boundary parameter $k$.  The dashed line is the
solution to \refeq{vaceq} in the interval $[3\pi,4\pi]$, which is the
value of $u_0=u(0)$ for the $n=2$ metastable vacuum.
}
	\label{boundaryOscillations}
\end{figure}

The Robin boundary for $k=0$ is the Neumann
limit, and indeed we find that the incoming antikink is perfectly
reflected into a kink without loss of energy, as shown in
\reffig{spaceTime}a.
For $k$ slightly above zero it is still possible for the antikink to
reflect into a kink, as shown in \reffig{spaceTime}b,
although some radiation is also generated, and
energy is also left at the boundary since it ends up in the metastable
$n=2$ vacuum. 
The region where this process occurs is labelled
I of \reffig{map}, and an approximation for its shape can be obtained
by noting that the final
state energy must be at least $E^{(2)}(k)+8$, where $E^{(2)}(k)$ is
the energy of the $n=2$ metastable vacuum as found in \refsec{Vac},
and $8$ is the lower bound on the energy of the final-state kink. The
initial energy is just that of an antikink with velocity $v_0$, which is
$8\gamma(v_0)$, where $\gamma(v_0)=(1-v_0^2)^{-1/2}$ is the Lorentz
factor. Some of this energy might be converted to radiation in the final
state, so region I must lie within the region
\begin{equation}
	8\gamma(v_0) \ge 8+ E^{(2)}(k).
\label{energybound}
\end{equation}
The boundary derived from this expression is the dashed line in
\reffig{map}. It is a good approximation to the true boundary of
region I while $v_0$ remains small, but
clearly diverges from it at higher values.
The reason is that after the initial collision,
boundary oscillations can be excited, which in turn produce
radiation.  For higher energies and larger values of $k$
this effect becomes significant and
the radiation produced is unaccounted for in our discussion.  Compare, for
example, the boundary oscillations for two different choices of $v_0$
and $k$ shown in
\reffig{boundaryOscillations}.

The subsequent behaviour as $k$ increases further depends on the value of
$v_0$.  If for the moment we restrict to 
$v_0 \lesssim 0.877$ then the first change, occuring as
the energy bound just discussed comes into play, is that
the reflected kink does not have enough 
energy to escape the boundary and
instead recollides with it and
reflects back as an antikink. This process is shown in
\reffig{spaceTime}c; it lies in region V$_a$.

\begin{figure}
\centering
\begin{subfigure}{1.0\linewidth}
\centering
\includegraphics[width = 0.8\linewidth]{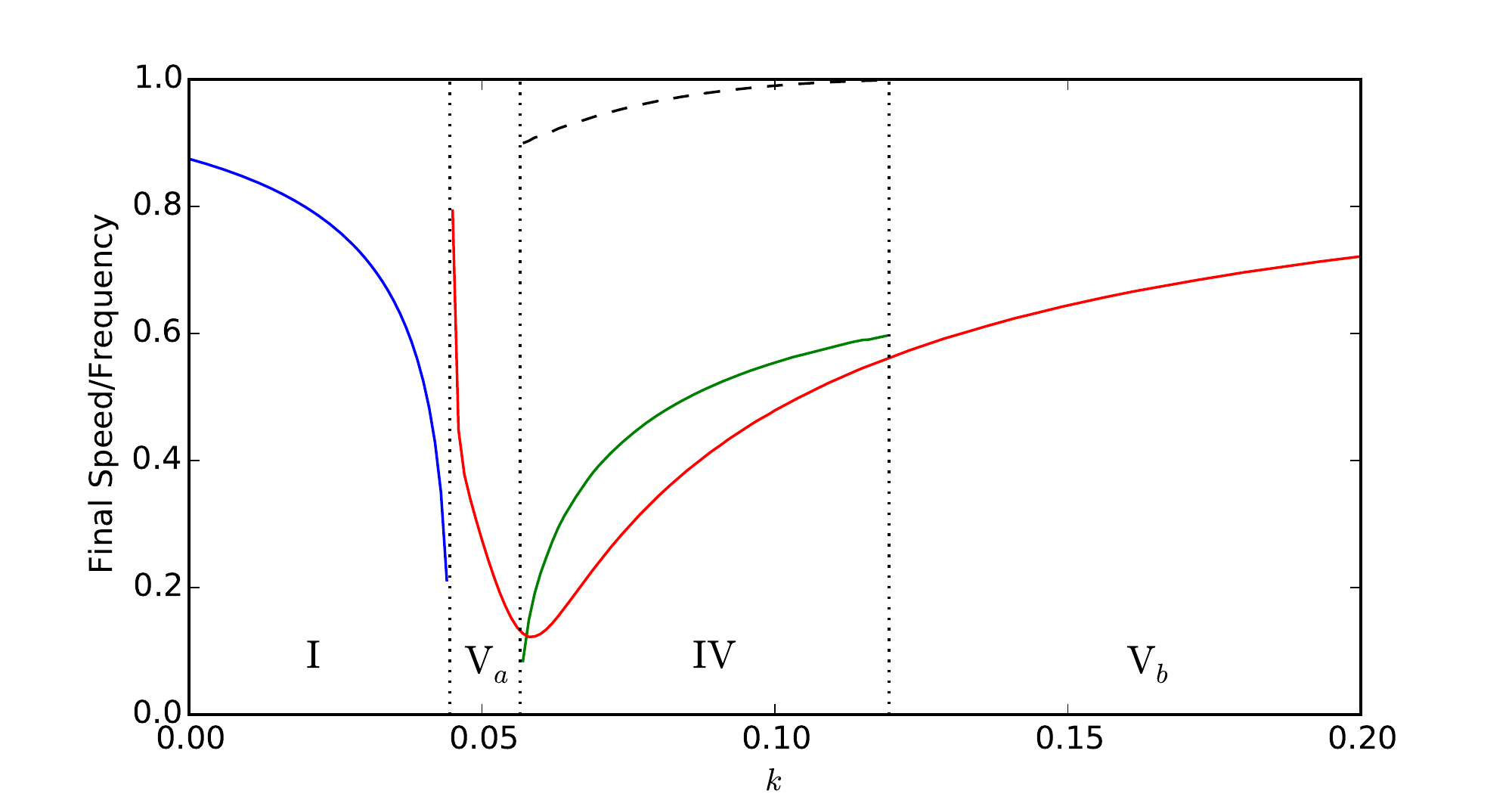}
\caption{\small $v_0=0.875$}
	\label{kinematicsMedium}
\end{subfigure}
\begin{subfigure}{1.0\linewidth}
\centering
\includegraphics[width =
0.8\linewidth]{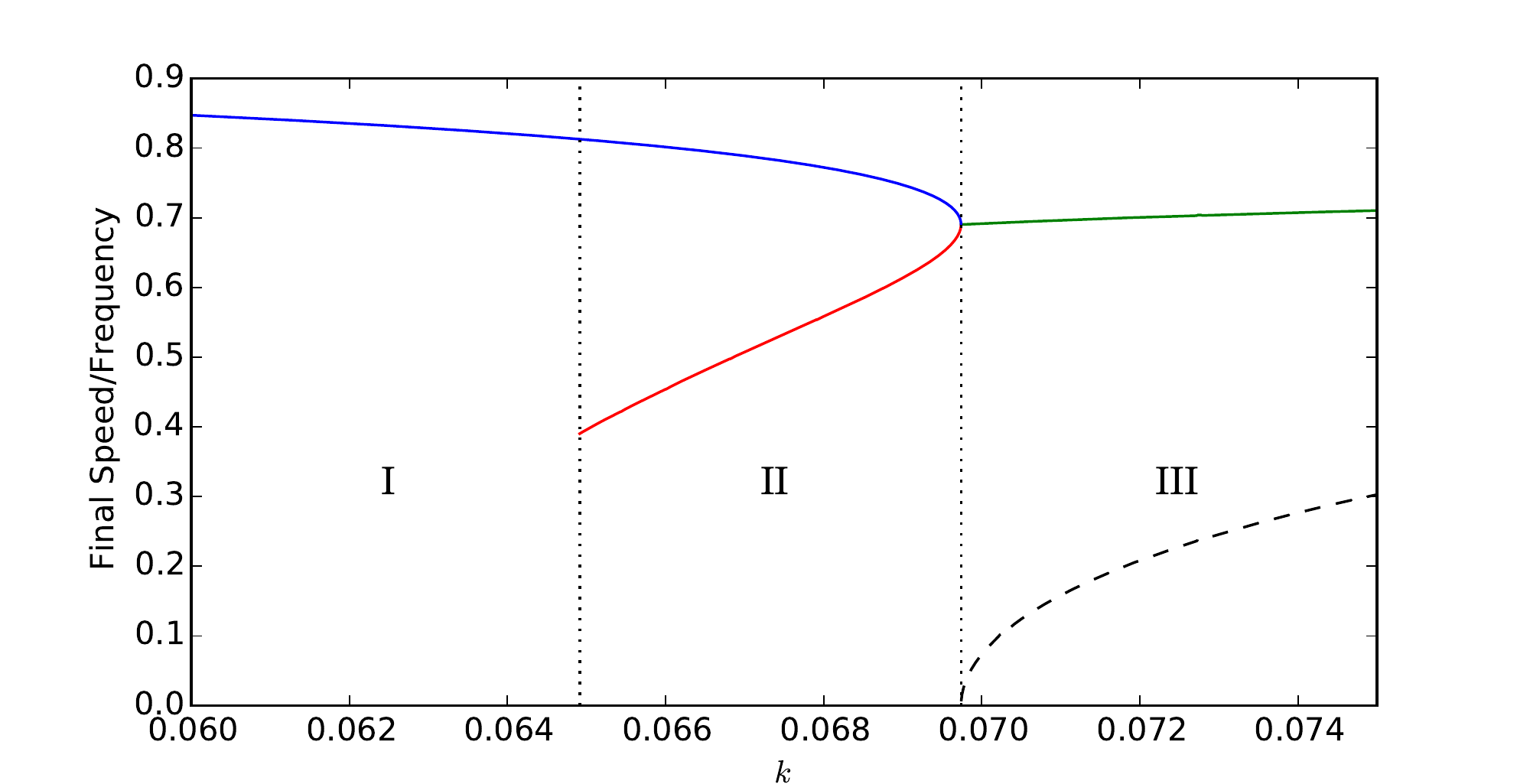}
\caption{\small $v_0=0.95$}
	\label{kinematicsHigh}
\end{subfigure}
\caption{\small
The outgoing kink speed
(blue), antikink speed (red), breather speed (green) and frequency
(black, dashed) after the collision of an antikink with initial
velocity (a) $v_0 = 0.875$ and (b) $v_0=0.95$ with the 
Robin boundary parameterised by $k$.  In each case only the speed
and frequency of the high energy breather, with $\omega < 0.999$, is
shown.}
	\label{kinematicsBoth}
\end{figure}

Increasing $k$ yet further brings us into
region IV, where in addition to an antikink and some radiation 
a relatively high-energy (low frequency) breather
is produced in the recollision, moving either slower than the
antikink (\reffig{spaceTime}d) or faster (\reffig{spaceTime}l).  
At the left
boundary of
region IV the breather speed goes to zero and indeed in parts of
region V$_a$ its presence can still be detected, trapped at the
boundary as in \reffig{spaceTime}e.  
This can be clearly seen in
\reffig{snapshotMess} where on the left hand edge of the plot (in V$_a$) the
boundary value of the field at the end of the simulation
oscillates a great deal, corresponding to the trapped breather,
while on the bottom right (in IV) it is
always very close to zero as the breather has escaped.  By contrast,
as the right boundary of region IV is approached,
the breather frequency goes to one and the breather energy goes
to zero, marking the transition to region V$_b$ where there is again just
an antikink in the final state. The whole sequence of transitions is
illustrated in \reffig{kinematicsMedium}.

For higher 
initial antikink velocities near to $1$ there is sufficient energy
in the initial state to produce both a kink and an antikink: this
occurs in region II, and
is illustrated in \reffig{spaceTime}f. The initial
collision produces a kink with enough energy being left at the
metastable boundary that some time later it decays with the emission of an
antikink.  As $k$ increases within region II,
the speeds of the kink and antikink approach the same value,
and the time between the release of the kink and antikink
becomes smaller. Ultimately the kink and antikink
`fuse' into a very low frequency (loosely bound) breather as shown in
\reffig{spaceTime}g; this marks the
transition from region II to region III, ending the sequence of
transitions shown in \reffig{kinematicsHigh}.
As $k$ increases further the frequency of this breather increases 
and its constituent kink and antikink become more tightly bound.  
We should also note that at the lower tip of
region III the high energy breather produced in the initial collision
can itself recollide with the Robin boundary, producing an extremely
complicated pattern of
results which we discuss in greater detail in
\refsec{BreatherRebound}.

To understand the transition from region III to region IV as $k$ 
increases even further, we note
that (just as was the case for the emission of a kink) when only
breathers are emitted the boundary is left in a metastable
vacuum, with the field suffering some
deformation near the boundary in order to satisfy
the boundary condition. With increasing $k$ the barrier to the decay
of this metastable vacuum decreases: in
\reffig{spaceTime}h (in region III)
it is still high enough that although an
antikink emerges it is unable to escape from
the boundary, while in \reffig{spaceTime}i (in region IV)
it does escape, the boundary relaxing to the true ($u=0$)
ground state.

As discussed above, increasing $k$ inside region VI increases 
the frequency of the emitted breather towards one.
Its energy correspondingly decreases to zero, whereupon it
disappears from the final state, leaving just a reflected antikink
as in \reffig{spaceTime}j. This marks the transition to the
Dirichlet-like behaviour of region V$_b$.

Finally we note that it is also possible to 
find final-state breathers at lower energies, in region VI
of \reffig{map}, although the mechanism is somewhat different than
that for region III.  
This process is shown in \reffig{spaceTime}k\,: after initially
rebounding the antikink fails to escape the boundary and instead forms
a breather upon colliding with the boundary a second time.
This breather appears to collide with the boundary multiple times and
may eventually escape the boundary, as in \reffig{spaceTime}k, or fail
to do so over the time we evolve the sine-Gordon equation.  This
behaviour can be traced to the phase dependence of breather/boundary
collisions and is discussed further at the end of
\refsec{BreatherRebound}.
In this region we also often detect several very low energy breathers
with frequency $\omega > 0.999$.

\section{Resonance structure}
\label{BreatherRebound}

\begin{figure}
	\includegraphics[width = \linewidth]{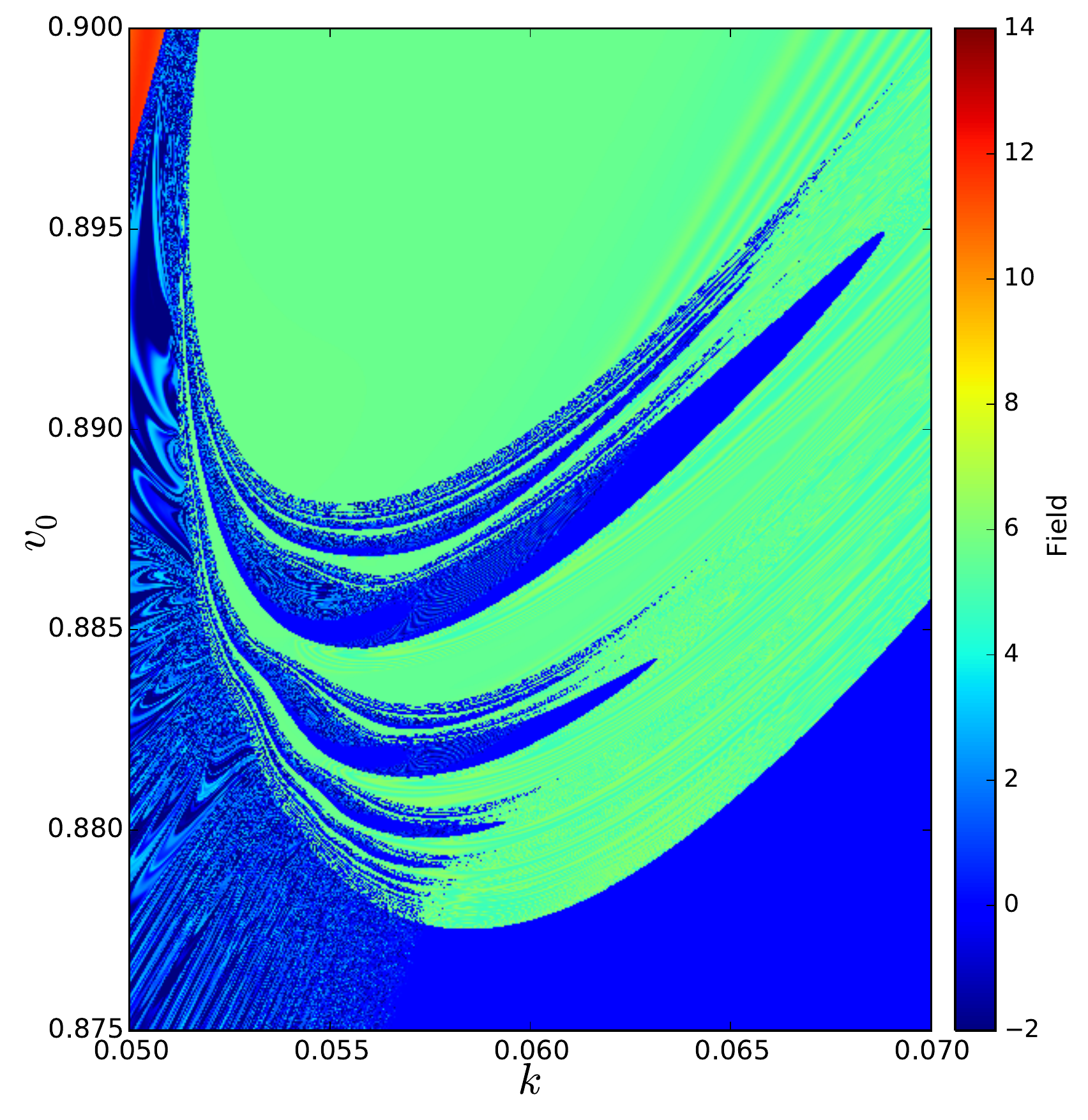}
	\caption{\small
A zoomed-in plot of the shaded area in \reffig{map},
showing the value of the field at $x=0$, $t=t_f=\abs{x_0}/v_0 + 1000$ for
an initial antikink with velocity $v_0$, position $x_0$, and boundary 
parameter
$k$.  The dark blue bands, where $u(0,t_f)$ is near zero, correspond to an antikink being
emitted, while in the light green areas, where $u(0,t_f)$ is near $2\pi$, only breathers are emitted.  In
between these areas are indeterminate regions where a very slight
change in the initial parameters can cause an antikink to be produced
or not.
The oscillations in the boundary value of the field on the
left of the plot are due to a breather becoming trapped at the boundary, only
decaying very slowly there, in contrast to behaviour on 
the bottom right where this breather is able to escape and the field
relaxes to zero much more quickly.
The line separating these two regions, running from approximately 
$k=0.0565$, $v_0=0.875$
to $k=0.0574$,
$v_0=0.8776$, is the top portion of
the boundary between regions V$_a$ and IV
in \reffig{map}.
	}
	\label{snapshotMess}
\end{figure}

\begin{figure}
	\includegraphics[width=\linewidth]{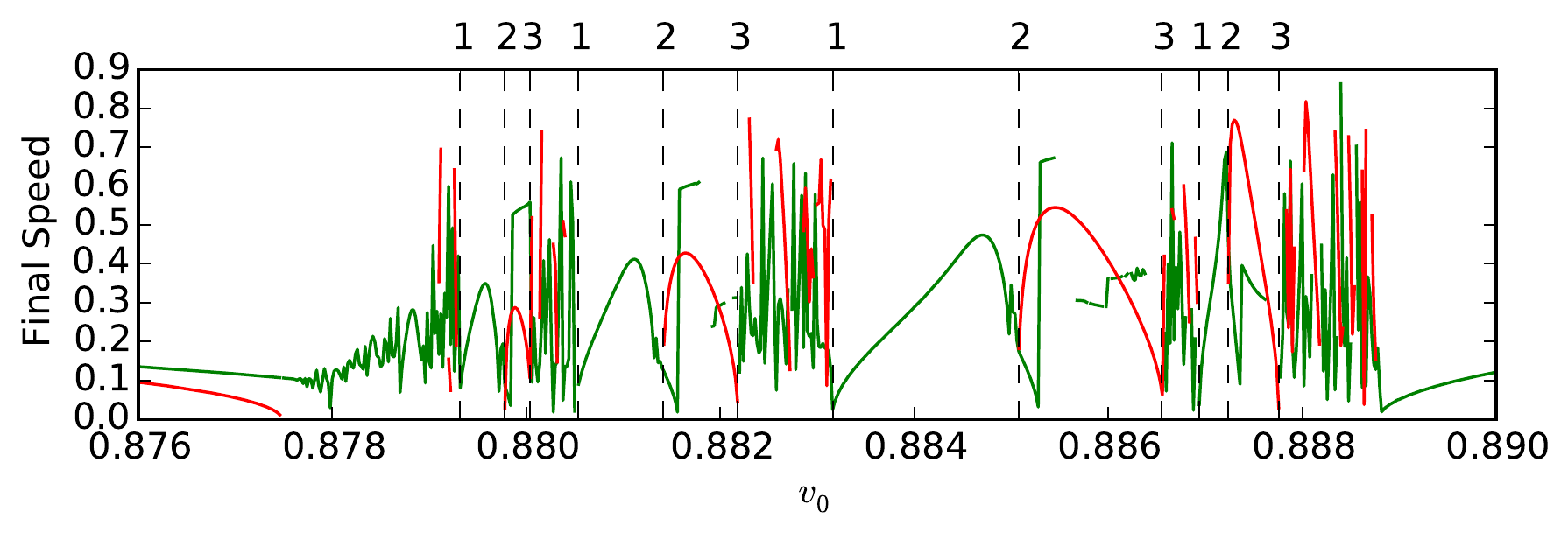}
	\caption{\small
The speed for the highest energy breathers (green) and antikinks (red)
produced by an antikink with initial velocity $v_0$ colliding with a
Robin boundary with $k=0.058$.  The bands shown in
\reffig{snapshotMess} correspond to the regions between the 1, 2, 3
labels.  Between 1 and 2 there is a resonance window for the
production of breathers, while between 2 and 3 there is an antikink
dominated resonance window and between 3 and 1 an indeterminate region
where a slight change in the initial parameters gives drastically
different results.}
	\label{kinematicsMess}
\end{figure}

\begin{figure}
\includegraphics[width=0.9\linewidth]{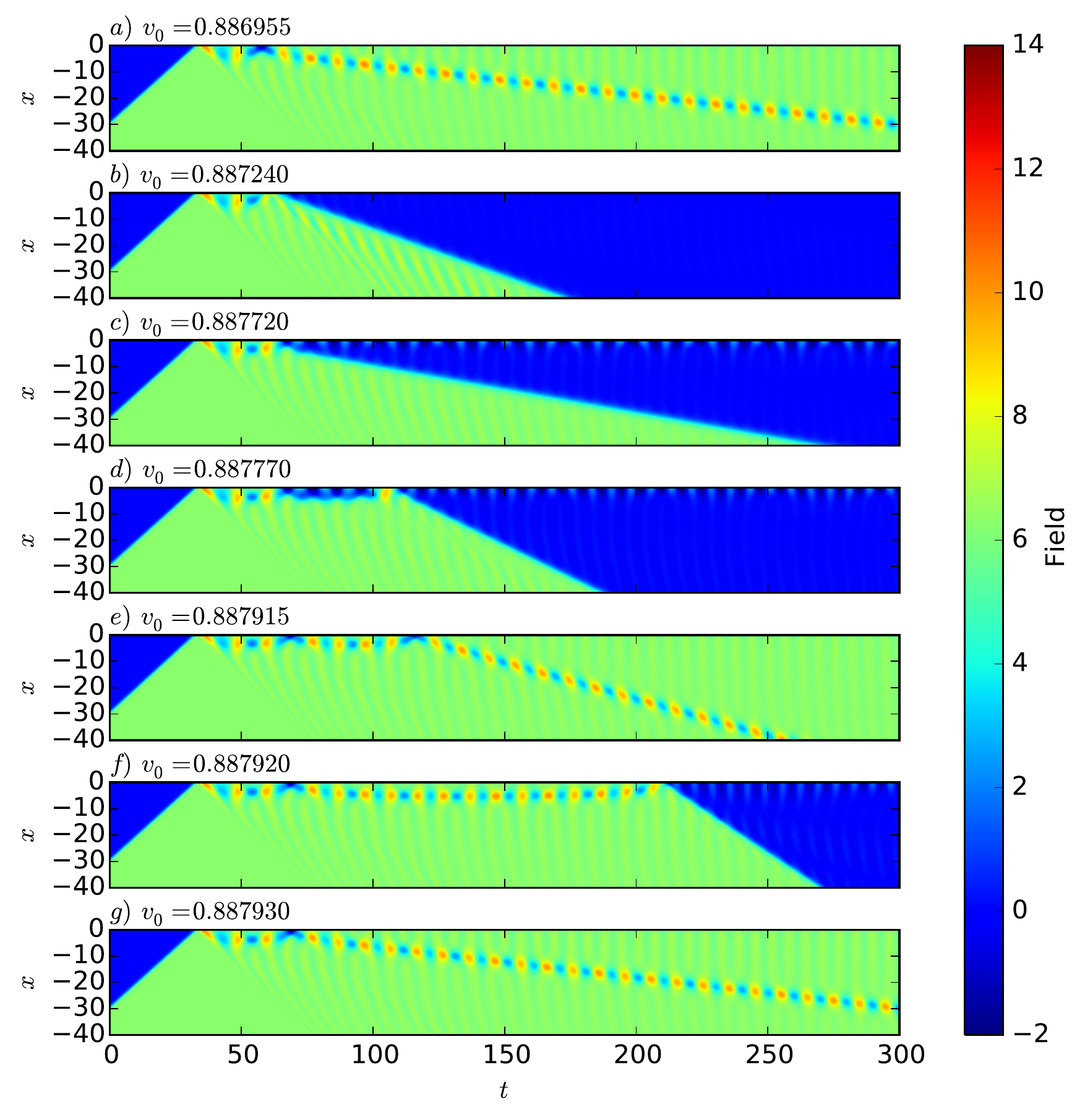}
	\caption{\small
Spacetime plots of an antikink with initial velocity $v_0$
colliding with a Robin boundary with boundary parameter $k=0.058$. For each
plot the soliton and breather content of the final state, excluding
breathers where $\omega>0.999$, is:
a) a breather;
b) an antikink and a breather;
c) an antikink;
d) an antikink;
e) a breather;
f) an antikink;
g) a breather. Note that in this regime the multiple recollisions
of excitations with the boundary cause the final state to depend very
sensitively both on the initial conditions and on any numerical errors
in the time evolution.
}
	\label{spaceTimeMess}
\end{figure}

Perhaps the most striking feature of our phase diagram is the
`chaotic' region shown in \reffig{snapshotMess}, which is 
reminiscent of the well-known patterns of resonance windows found in
the non-integrable $\phi^4$ theory 
\cite{Campbell:1983xu,Anninos:1991un,Goodman:2007}. 
A window-like pattern of final velocities can also be seen in
\reffig{kinematicsMess},
a cross-section of this region at $k=0.058$. Naively this might be
surprising, as sine-Gordon kinks and antikinks lack the internal mode
responsible for the resonance windows of the $\phi^4$ theory on the
full line \cite{Campbell:1983xu}. 
The critical distinction in the presence of a non-integrable boundary
is that the initial antikink
collision can create an intermediate breather, which \textit{does} have
an internal mode, and which furthermore can be attracted back
towards the boundary to collide with it
again.  Several examples of this process, with dramatically different
final states, are shown in \reffig{spaceTimeMess}.

\begin{figure}
\includegraphics[width=\linewidth]{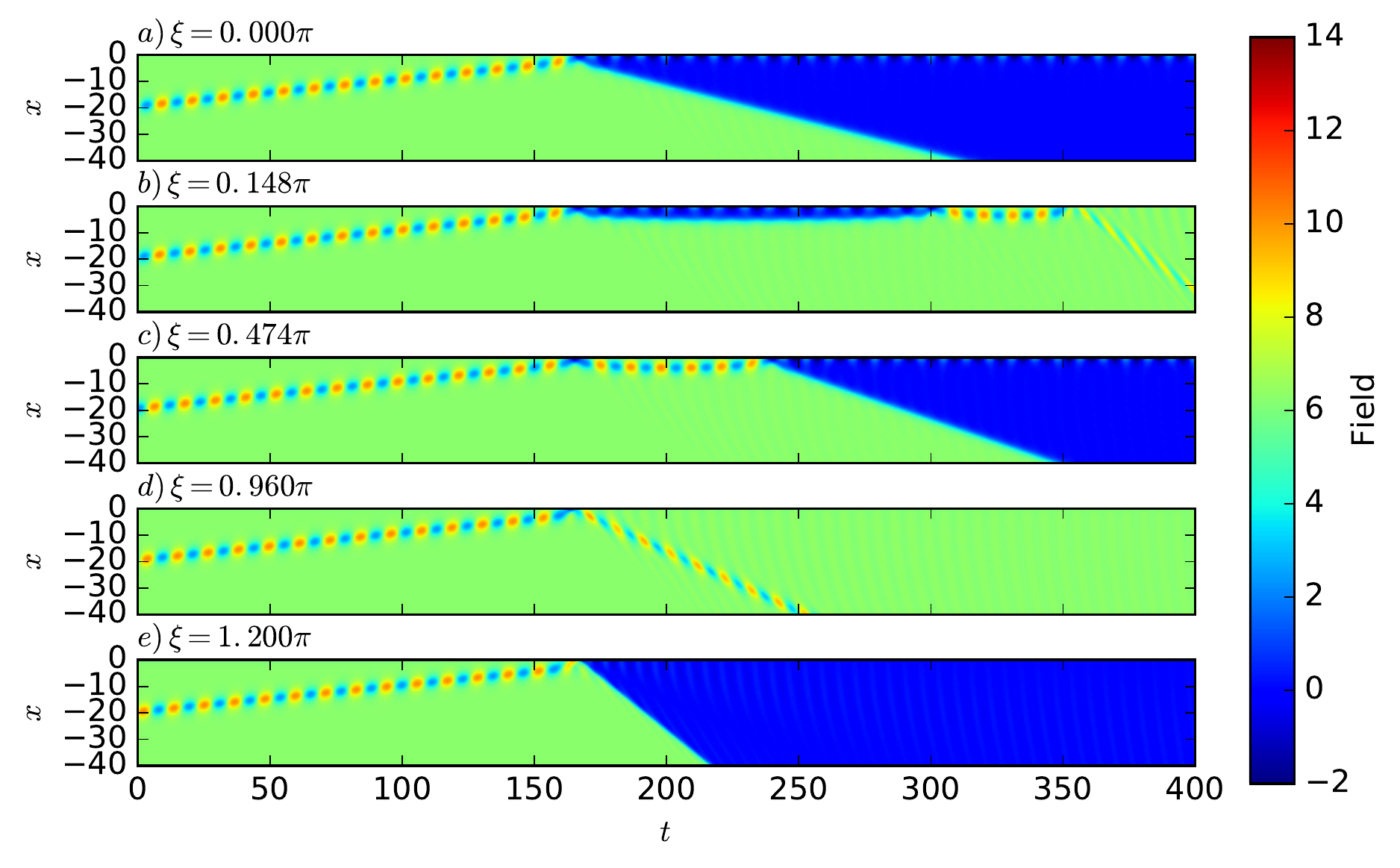}
\caption{\small
Spacetime plots showing a breather with initial velocity $v_0=0.1$,
frequency $\omega=0.55$ and a variety of initial phases
$\xi \in [0,2\pi)$ colliding
with an $n=1$
 metastable Robin boundary with boundary parameter $k=0.058$. An antikink
has been placed at $x=1.79$ in order to satisfy the boundary
condition
and model the environment in which an intermediate breather created
by an antikink collision would recollide with the boundary. 
In each case the antikinks and/or breathers escaping from the boundary
are:
a) an antikink;
b) a breather;
c) an antikink;
d) a breather;
e) an antikink and a breather.
}
\label{spaceTimePhase}
\end{figure}

The reason for the variety of outcomes in \reffig{spaceTimeMess} is that
the result of a breather colliding with a metastable Robin boundary is 
highly dependent on the point in the breather's cycle at which it
hits the boundary.  As shown in \reffig{spaceTimePhase}, simply
changing the initial phase of the breather can create a completely
different final state after collision with the boundary.  In
\reffig{spaceTimePhase}a the breather fissions into an antikink
and a boundary breather, while \reffig{spaceTimePhase}c has a similar
outcome but only after an intermediate breather is created and
recollides with the boundary.  \reffig{spaceTimePhase}d shows the
breather being reconfigured into a breather of lower mass and higher
speed, and \reffig{spaceTimePhase}b shows this outcome happening via
an intermediate antikink and breather.

This strong phase dependence suggests that the breather and antikink
resonance windows exemplified in \reffig{kinematicsMess} occur when
the frequency and initial phase and velocity
of the intermediate breather are such
that it recollides with the boundary at exactly
the `right' phase to produce an antikink and/or
breather which escapes the boundary.  Of course for our model it is
the initial antikink speed, $v_0$, and the boundary parameter, $k$,
which indirectly controls all the characteristics of the intermediate
breather.

Let us consider moving between these resonance windows in more detail
for the specific case where $k=0.058$, as shown in
\reffig{kinematicsMess}.  Starting at a label 1 in
\reffig{kinematicsMess} the intermediate breather collides with the
boundary and produces a breather which then escapes the boundary, as
in \reffig{spaceTimeMess}a.  As $v_0$ increases first an antikink is
produced in addition to the breather, shown in
\reffig{spaceTimeMess}b and corresponding to label 2 in
\reffig{kinematicsMess}.  Then as the breather speed decreases it
becomes trapped at the boundary as in \reffig{spaceTimeMess}c.
Eventually the antikink fails to escape the boundary, which marks
label 3 in \reffig{kinematicsMess} and the beginning of the
indeterminate region which, as we see in \reffig{spaceTimeMess}d, e
and f is due to multiple intermediate antikinks and breathers
scattering off the boundary.  Of course each successive intermediate
breather produced will have its own resonance windows that may allow a
breather or antikink to escape.  So the final result we see for the
indeterminate regions in \reffig{kinematicsMess} is a product of one or
more of these nested resonance windows, giving these regions their
chaotic appearance.  As $v_0$ continues to increase there will
eventually come a point where the phase of the intermediate breather
as it collides with the boundary cycles back to its original value
where a breather is produced.  For example, in
\reffig{spaceTimeMess}g the breather has undergone an additional full
cycle in its oscillation compared to \reffig{spaceTimeMess}a and
the final states are quite similar. 

For a sufficiently high $v_0$ (the precise value being dependent upon
$k$) the breather formed after the initial collision has enough energy
to escape the boundary in the first instance, as in the rest of region
III in \reffig{map}.

For sufficiently low $v_0$ (again, depending on $k$) an antikink
with (in region IV) or without (in region V) a breather is produced.
Comparing \reffig{spaceTime}l and \reffig{spaceTime}e to
\reffig{spaceTimeMess}b and \reffig{spaceTimeMess}c suggests that
this transition to regions IV and V can be interpreted as the
intermediate breather becoming very short lived and colliding with the
boundary before oscillating a full cycle.  Because the breather is so
short-lived it appears very much like a short-lived kink.  This
coincides with the interpretation discussed in \refsec{RobinResults}
that as $k$ is increased from region I there is an intermediate kink
which exists for a progressively shorter time.  For example, compare
the progression from \reffig{spaceTime}c to d to e.

Finally, a basic explanation for the behaviour observed in region VI
and exemplified in \reffig{spaceTime}k, where the recollision of an
intermediate antikink creates a breather which may collide with the
boundary multiple times before escaping, is now apparent.  The
breather will only escape when its phase as it hits the boundary is
such that after the collision it has a mass and speed that allows it
to escape the boundary, schematically similar to the case shown in
\reffig{spaceTimePhase}d. We therefore expect this region to exhibit a
similarly chaotic pattern of breather escapes as was seen in the lower
portion of region III for antikink escapes.
Note though that the total energy available available to the breather
is less than the escape energy of an
antikink, since the breather itself was formed by a returning 
antikink. For this reason any chaotic patterns will only be visible in
the breather spectrum, making them much harder to see than in region
III. Further (and higher-precision) study will be required before the
full picture in this region is clear.

\section{The Robin boundary with $k <0$}

Here we will make some brief remarks on the case when $k<0$.
It is known that the integrable boundary, \refeq{integrableBC}, with $\widehat u = 0$ is 
unstable for ${K \leq -1/2}$ since the boundary potential,
${8K(1-\cos(u/2))}$, then allows for a family of degenerate zero energy
solutions \cite{Fujii1995}.  For example, with an
initial condition of $u=0$, a kink may be emitted without any loss of
energy. 

The Robin boundary appears to exhibit similar instabilities.  For
${-0.051 \lesssim k <0}$ the incoming antikink is converted into a
kink but we observed that for ${k \lesssim -0.051}$  additional kinks
are produced.  This threshold can be approximated by noting that if
${k \lesssim -0.045}$ then ${k(6\pi)^2 - k(4\pi)^2 \geq -8}$ so that
the energy required to release the additional kink is offset by the
energy released due to the boundary moving up ${2\pi}$.  If this
inequality is satisfied then so will ${k(2(n+1)\pi)^2 - k(2n\pi)^2
\geq -8}$ for ${n > 2}$ which allows for an infinite number of kinks
to be produced from the boundary.

As k becomes increasingly negative the numerical simulation becomes
unstable with even the slightest increase in the value of the field at
the boundary from its initial value of zero causing the field at the
boundary to blow up to infinity.

\section{Conclusions}
\label{conclusions}

We have examined the wide range of processes and outcomes
arising from the collision of an antikink solution to the sine-Gordon
equation with a non-integrable Robin boundary.
An important feature of our analysis was the numerical implementation
of the direct scattering transform which enabled us to disentangle the
excitation content of the final state in an efficient manner.
Even though integrability is only broken at one point, the structure
turned out to be very rich:
Figs.\
\ref{snapshotFull} and
\ref{maps} summarise the broad features, while Figs.\ \ref{snapshotMess}
and \ref{kinematicsMess} reveal a complicated
finer structure.
In the integrable Neumann and Dirichlet limits 
the results of the collisions
approach the exact solutions for these boundaries:  close to
$k=0$ the antikink reflects into a kink (region I), while for large $k$
the antikink retains its character as an antikink (region V$_b$).
Away from these limits the non-integrability of the boundary allows
the production of a kink and an antikink (region II), high energy
breathers (region III), an antikink accompanied by a breather (region
IV), or the annihilation of the initial antikink into either radiation or
low energy breathers (region VI).
The most exotic features observed were the resonance structures of
\reffig{snapshotMess}, and their origin was traced to the phase dependence
of the recollision of intermediate breathers with the boundary.

While an approximation to the boundary of region I was given, our
discussion was largely phenomenological and we have
not found analytical arguments for the shapes of the other regions.
Progress in this area would appear to require a greater
quantitative understanding of how the antikink initially collides with
the boundary.
For example, deriving the shape of region VI would require a model of
how much energy the incoming antikink loses in its initial collision
with the boundary for a given $v_0$ and $k$.  This would determine
whether the antikink has sufficient energy to escape the boundary.
In the case of the $\phi^4$ model on the full line a similar style of
analysis has yielded considerable insights
\cite{Campbell:1983xu,Anninos:1991un,Goodman:2007}, so this
should be a promising avenue for further work.

Perhaps the most interesting result was the intricate resonance
structure seen in \reffig{snapshotMess}, which we traced to
the phase dependence of the
recollision of intermediate breathers with the boundary.
This behaviour is clearly deserving of a more detailed analysis.
Such resonance phenomena greatly increase sensitivity not
only to initial conditions but also to numerical error, and while we
tried to keep these issues under control by varying the time and space
steps in our simulations, a closer examination of the patterns of
resonance windows using more sophisticated numerical
methods would be very valuable,
both in the regions of windowed antikink escape shown in
\reffig{snapshotMess}, and in also in the portions of
region VI which appear to have a similar pattern of escape
and non-escape, but for breathers rather than antikinks. Work on this
question is in progress.
To make further analytical headway, a better understanding of the way
that the initial antikink velocity combines with
the boundary parameter 
to determine the characteristics of the first intermediate breather,
and how these in turn affect its subsequent recollision with the
boundary, will be required.  A first step is therefore
likely to involve a more-detailed and higher-precision study of how
a breather with a given initial phase,
frequency and velocity collides with the various 
metastable Robin boundaries.

It would also be interesting to see whether the fact that the model
remains integrable away from the boundary can be exploited in a more
direct way, possibly within the framework of the Fokas (or unified)
method.  With respect to integrable PDEs on the half line this 
can be viewed as a generalisation of the inverse scattering transform
\cite{Pelloni2015,Fokas2008}.  

Specialised to the sine-Gordon equation, the Fokas method requires not
only the initial data $u(x,0)$ and $u_t(x,0)$, but also the, most likely
unknown, boundary data $u(0,t)$ and $u_x(0,t)$.
A key component in this method is therefore the so-called `global
relation', an equation relating the spectral functions associated with
the initial data and the boundary data.
If one considers a boundary problem where $u(0,t)$ is a known function
of time then it is possible to derive a `Dirichlet to Neumann map' and
obtain a perturbative expansion for the unknown $u_x(0,t)$
\cite{Fokas2005,Hwang2014}.  An analogous procedure can be carried out
when $u_x(0,t)$ is known and $u(0,t)$ unknown.
Alternatively, for certain boundary conditions termed `linearizable'
there is an additional symmetry of the Lax pair eigenfunction which
makes it possible to solve the global relation algebraically,
bypassing the need for a perturbative solution for $u_x(0,t)$.
For sine-Gordon the known linearizable boundary conditions are nothing
but the integrable boundaries of \refeq{integrableBC} 
\cite{Fokas2003}.

However, the sine-Gordon equation with a Robin boundary of the type we
have considered does not fit into either of these cases.  That is to
say it is not linearizable and we do not know \textit{a priori}
$u(0,t)$ or $u_x(0,t)$ for $t>0$.  At present we are unaware of a
scheme that would allow us to apply the Fokas method given only a
relationship between $u(0,t)$ and $u_x(0,t)$ such as, $u_x(0,t) + 2k
u(0,t)= 0$. Given the complexity of the behaviour that we have
observed, this would seem to present an interesting challenge for the
wider applicability of the method.

Another direction for future work is to investigate whether the
classical phenomena that we have found have their counterparts in the
corresponding boundary quantum field theories. There has been a certain
amount of work treating non-integrable \textit{bulk} quantum field
theories as deformations of integrable theories (see for example
\cite{DMS1996}), and it would be interesting to apply similar ideas to
problems where integrability is instead broken just
at a boundary. In particular, since
all excitations are asymptotically far from that boundary,
one would expect the space of in- and out- states to be
exactly the same as for the integrable theory. This major
simplification should enable significantly more progress to
be made than in previously-studied bulk examples.

Finally, and returning to classical considerations, 
we note that the method for numerically obtaining the soliton and
breather
content after collision is not limited to the sine-Gordon equation and
could easily be extended to any integrable model that permits
a solution via the inverse scattering transform with any
non-integrable boundary or even defect.  The analysis of other
boundary or defect models may well benefit from the additional
information this provides.

\section*{Acknowledgements}

\vspace {-7pt}

We wish to thank Vincent Caudrellier and Ed Corrigan for helpful
comments and encouragement during the later stages of this work, and
Giuseppe Mussardo for useful correspondence. PED
would also like to thank Tomasz Romanczukiewicz and Yasha Shnir for
collaboration on related topics.
PED was supported in part by an STFC
Consolidated Grant, ST/L000407/1, and also
by the GATIS Marie Curie FP7 network (gatis.desy.eu)
under REA Grant Agreement No 317089.
RP was supported by an EPSRC vacation bursary
while the majority of this research was carried out and gratefully
acknowledges the support of an EPSRC studentship at present.

\vspace {-2pt}

\section*{References}
\raggedright

\end{document}